\documentclass[a4paper,11pt]{article}
\pdfoutput=1

\usepackage{jcappub}
\usepackage[T1]{fontenc}
\usepackage{textcomp}
\usepackage{gensymb}
\usepackage{array}
\usepackage{fixltx2e}
\usepackage{graphicx}
\usepackage[english]{babel}
\usepackage{blindtext}
\usepackage{lipsum}
\usepackage{color}
\usepackage{amsmath}
\usepackage{amssymb}
\usepackage{longtable}
\usepackage{multirow}
\usepackage{array}
\usepackage{bmpsize}
\usepackage{rotating}
\usepackage{ifthen}
\usepackage{fixltx2e}
\usepackage{lscape}
\usepackage{booktabs}
\usepackage{float}
\usepackage[framemethod=tikz]{mdframed}

\newcommand{\new}[1]{{\textcolor[rgb]{0, 0, 0}{ #1}}}
\newcommand{\newb}[1]{{\textcolor[rgb]{0, 0, 0}{ #1}}}
\newcommand{\newc}[1]{{\textcolor[rgb]{0, 0, 0}{ #1}}}
\newcommand{\newd}[1]{{\textcolor[rgb]{0, 0, 0}{ #1}}}
\newcommand{\newe}[1]{{\textcolor[rgb]{0, 0., 0}{ #1}}}
\usepackage{tcolorbox}

\DeclareGraphicsExtensions{.jpg, .png, .pdf, .eps}

\title{QUBIC VII: The feedhorn-switch system of the technological demonstrator}

\author[1,2]{F.~Cavaliere}
\author[1,2]{A.~Mennella}
\author[3,4]{M.~Zannoni}
\author[5]{P.~Battaglia}
\author[6,7]{E.S.~Battistelli}
\author[6,7]{P.~de~Bernardis}
\author[8]{D.~Burke}
\author[6,7]{G.~D'Alessandro}
\author[6,7]{M.~De~Petris}
\author[1,2]{C.~Franceschet}
% ...

\affiliation[1]{Universit\`a degli studi di Milano, Milano, Italy}
\affiliation[2]{INFN sezione di Milano, 20133 Milano, Italy}
\affiliation[3]{Universit\a`{a} di Milano - Bicocca, Milano, Italy}
\affiliation[4]{INFN sezione di Milano - Bicocca, 20216 Milano, Italy}
\affiliation[5]{INAF/OAS Milano Section}
\affiliation[6]{Universit\a`{a} di Roma - La Sapienza, Roma, Italy}
\affiliation[7]{INFN sezione di Roma, 00185 Roma, Italy}
\affiliation[8]{National University of Ireland, Maynooth, Ireland}
% ...

\emailAdd{aniello.mennella@fisica.unimi.it, mario.zannoni@unimib.it}

\abstract{We present the design, manufacturing and performance of the horn-switch system developed for the technological demonstrator of QUBIC (the $Q$\&$U$ Bolometric Interferometer for Cosmology). This system \new{consists} of 64 back-to-back dual-band (150\,GHz and 220\,GHz) corrugated feed-horns \new{interposed} with mechanical switches used to select desired baselines during the instrument self-calibration. We manufactured the horns in aluminum platelets milled by photo-chemical etching and mechanically tightened with screws. The switches are based on steel blades that open and close the \newb{waveguide} between the back-to-back horns and are operated by miniaturized electromagnets. \newd{The measured electromagnetic performance of the feedhorns agrees with simulations. In particular we obtained a return loss around $-$20\,dB up to 230\,GHz and beam patterns in agreement with single-mode simulations down to $-$30\,dB. The switches for this prototype were designed and built for the 150\,GHz band. In this frequency range we find return and insertion losses \newb{consistent} with expectations ($<-$25\,dB and $\sim-0.1$\,dB, respectively) and an isolation larger than 70\,dB.} \newd{In this paper we also show the current development status of the feedhorn-switch system for the QUBIC full instrument, based on an array of 400 horn-switch assemblies.}}


\begin{document}
\maketitle
\flushbottom  

\section{Introduction}
\label{sec_introduction}

	This paper is part of a set describing the current status of QUBIC (Q and U Bolometric Instrument for Cosmology), an experiment based on the concept of bolometric interferometry \newc{\cite{2011APh....34..705Q, 2019Univ....5...42M}} \new{that will observe the sky at 150\,GHz and 220\,GHz from the Argentinean Alto Chorrillo site and is designed to tightly constrain} the $B$-mode polarization anisotropies of the Cosmic Microwave Background (CMB). 

\newc{In the landscape of CMB polarization experiments, QUBIC is currently the only instrument based on this concept that combines the sensitivity of transition edge sensor (TES) bolometers with the advantages of interferometers. The combination gives features including self-calibration, in-band spectral resolution, which we call spectral imaging \cite{2020.QUBIC.PAPER1, 2020.QUBIC.PAPER2}, and interferometric rather than photometric instrumental systematic effects.} In this paper we describe the design, manufacturing and testing of the horn-switch array developed for the QUBIC technological demonstrator (TD), a prototype instrument that will demonstrate the application of bolometric interferometry to CMB polarization measurements with dedicated sky observations carried out from the QUBIC site.


\newc{As shown in figure~\ref{fig_intrument_system_overview}, each horn in the array is connected to its twin with a straight circular waveguide section that can be kept open or closed by mean of a mechanical switch.} When the instrument is in calibration mode it observes an artificial source in the far field and we can close subsets of horn pairs to select only certain baselines. This particular calibration scheme, that we call \textit{self-calibration}, allows us to determine the instrument parameters with high precision thanks to the redundancy provided by the large number of available baselines. The interested reader can find all the details of the self-calibration concept applied to QUBIC in \newd{Bigot-Sazy et al. \cite{Bigot-Sazy2013}}.

\new{The QUBIC TD horn-switch system has 64$+$64 back-to-back horns and 64 switches, while in the QUBIC full instrument (FI) this number is raised to 400 (i.e., 400$+$400 horns and 400 switches). In this paper we describe in detail the manufacturing and testing of the horn-switch system of the QUBIC TD and present briefly the advances of the development of the array that will be installed in the QUBIC FI.}

\new{The main challenge in this work has been to realize large arrays of feed-horns and switches with a cost-effective and scalable approach. Concerning feed-horns, we have found the optimal solution in the platelet technique combined with chemical etching. This technique, first introduced by Haas et al. in 1993 \cite{haas1993,haas1995}, allows one to build feed-horns by stacking metal plates that are drilled either by mechanical tools or by chemical etching. \newd{Several examples} of the application of the platelet technique in the field of CMB polarization experiments can be found in \newd{the literature \cite{bock2009, britton2010, deltorto2011,deltorto2015, 2012JLTP..167..917N, bischoff2013,mandelli2021}}. The waveguide switches are realized by insertion of a metallic blade in the guide connecting the back-to-back horns}. \newc{The preliminary challenge is presented by realizing a switch mechanism capable of moving the shutters at cryogenic temperature in a reliable manner.} Since we did not find similar devices described in the literature, we carried out a completely new research and development effort. 

In our paper we start with a top-level description of the system in section~\ref{sec_horns_switches_system}, then we present the back-to-back horn system in section~\ref{sec_back2back_prototype} and the switch array in section~\ref{sec_switches}. In each section we review the main requirements, discuss the manufacturing and testing techniques, present the testing results and provide an overview of the development of the arrays for the QUBIC FI. Finally, section~\ref{sec_conclusions} summarizes the conclusions and future prospects of our work.

%
\section{The QUBIC Instrument}
\label{sec_instrument_description}

\new{QUBIC employs an optical system consisting of back-to-back horns that select the relevant baselines and an optical combiner focusing on a bolometric
focal plane. The optical combiner forms interference fringes while the bolometers average their powers over timescales much larger than the period of the light waves. This is therefore the optical equivalent of a wide-band correlator in classical interferometry. The instrument operates at cryogenic temperatures thanks to a large cryostat described in \newd{Masi et al. \cite{2020.QUBIC.PAPER5}}.}


\begin{figure}[H]
    \begin{center}
        \includegraphics[width=12.5cm]{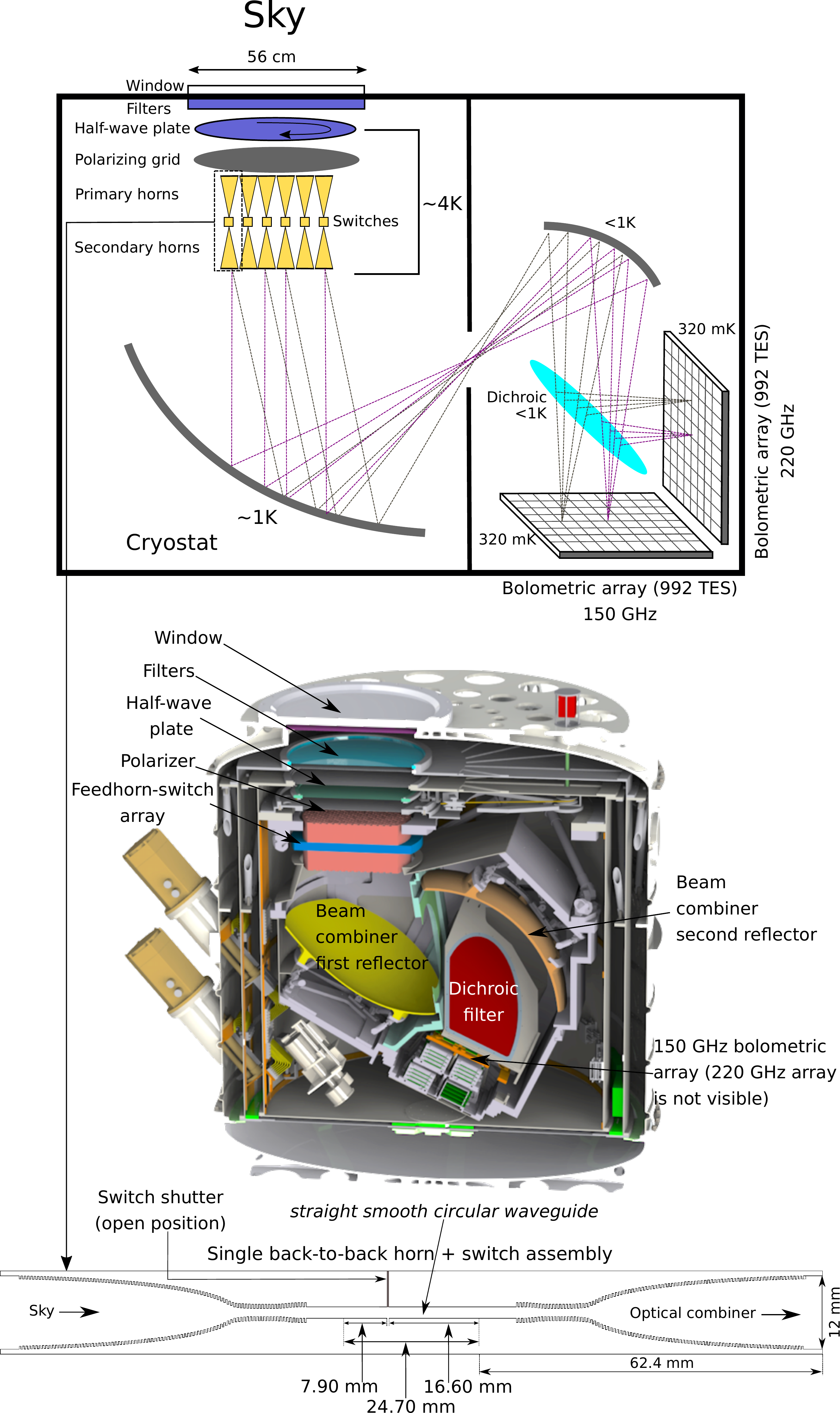}
    \end{center}
    \caption{\label{fig_intrument_system_overview}\newc{\textit{Top}: schematic of the QUBIC instrument. \newb{The filters in front of the half-wave plate are infrared filters described in table~1 of \newd{Masi et al. \cite{2020.QUBIC.PAPER5}}}. The cold stop is the septum between the two mirrors. \textit{Middle}: sectional cut of the cryostat showing the same sub-systems in their actual accommodation. \textit{Bottom}: detailed schematic of a pair of back-to-back horns connected by a circular \newb{waveguide} that can be shut by the switch blade (shown by the gray rectangle). Notice that the blade is not placed in the middle of the \newb{waveguide} section. The reason lies in the mechanical layout of the switch system (see section~\ref{sec_switches_requirements_design}, figure~\ref{fig_single_switch_elements}).} }
\end{figure}

\newlength{\colone}
\settowidth{\colone}{Synthesized beam FWHM [degrees]~~~}
\newlength{\coltwo}
\settowidth{\coltwo}{[131-169]~GHz}
\newlength{\colthree}
\settowidth{\colthree}{0.39 (150~GHz), 0.27 (220~GHz)}

\begin{table}[t]
    \renewcommand{\arraystretch}{1.}
    \begin{center}
        \caption{\label{table:qubic_params}\new{QUBIC main parameters}}
\begin{tabular}{p{\colone} p{\coltwo} p{5.5cm}}
          \hline
            Parameter& QUBIC-TD & QUBIC-FI \\
            \hline
            \hline
            Frequency channels \dotfill &150 GHz & 150 GHz \& 220 GHz\\
            Frequency range 150 GHz \dotfill &[131-169] GHz &[131-169] GHz\\
            Frequency range 220 GHz \dotfill &- &[192.5-247.5] GHz\\
            Window Aperture [m]\dotfill & 0.56 & 0.56 \\
            Number of horns\dotfill &64 &400\\
            Number of detectors\dotfill &248 &992$\times$2\\
            Detector noise [$\mathrm{W/\sqrt{Hz}}$]\dotfill & 2.05$\times 10^{-16}$ & 4.7$\times 10^{-17}$ \\
            Focal plane temp. [mK]\dotfill &300 &300\\
            Sky Coverage\dotfill &1.5\% &1.5\%\\
            Synthesized beam FWHM [degrees]\dotfill &0.68 &0.39 (150~GHz), 0.27 (220~GHz)\\
        \end{tabular}
        
    \end{center}
\end{table}

\newc{A schematic of the design of QUBIC is shown in figure~\ref{fig_intrument_system_overview} and the main instrument parameters are listed in table~\ref{table:qubic_params}. The sky signal ﬁrst goes through a 56 cm diameter, 25 mm thick window made of Ultra-High-Molecular-Weight Polyethylene followed by a series of thermal blocking ﬁlters, which are used to limit the infrared loading on the cryo-system. For a detailed description of the infrared ﬁlter stack in the cryostat, see table~1 in \newd{Masi et al.} \cite{2020.QUBIC.PAPER5}.}

\new{The next optical component is the
stepped rotating HWP which modulates incoming
polarization \newc{\cite{2020.QUBIC.PAPER6}}. A single polarization is then selected by the polarizing grid. \newc{This scheme makes the instrument largely insensitive to spurious polarization introduced by optical elements after the grid. A non ideal behavior of the system components could still introduce spurious polarization even after the polarizing grid (for example a fraction of the signal reflected by the polarizing grid could be reflected back in the main optical path, or the detectors could display some sensitivity to polarization). Laboratory measurements, however, have shown that the upper limit in spurious polarization is of the order of 0.4\% (\newc{see figure 13 in \newd{Torchinsky et al.} \cite{2020.QUBIC.PAPER3}}) and that this limit is uniform over the detector focal plane (\newc{section 9.2 in \newd{D'Alessandro et al.} \cite{2020.QUBIC.PAPER6}}).}}

\new{The next optical device is an array of 400~back-to-back corrugated horns made of an assembly of two 400-horn arrays, composed of 175~aluminium platelets (0.3~mm thick) chemically etched to reproduce the corrugations required for the horns to achieve the required
performance. An array of mechanical shutters (RF switches) separates the two back-to-back horn arrays in order to be able to close or open horns for self-calibration \newc{\cite{2020arXiv200810056T}}. The shutters are spring loaded and activated by applying a voltage to an induction coil.  As a result, the shutter requires continuous electrical current in order to remain closed.  To prevent heating of the 1~K horn array by the thermal dissipation by the shutters, no more than two switches may be simultaneously activated. However, due to the modularity of the switch electronics, a maximum of two switch per module can be operated at the same time. Both front and back horns are identical with a field of view of 13~degrees FWHM with secondary lobes below $-$25~dB.}

\new{As described in \newd{O'Sullivan et al.} \newc{\cite{2020.QUBIC.PAPER8}}, the back-horns directly illuminate the two-mirror off-axis Gregorian optical combiner which focuses the signal onto the two perpendicular focal planes.  A dichroic filter splits the incoming waves into two wide bands centred at 150~GHz for the on-axis focal plane and 220~GHz for the off-axis focal plane. The focal planes are each equipped with 992~NbSi Transition-Edge-Sensors cooled down to 320~mK using a sorption fridge; for a detailed description see \newd{Piat et al.} \newc{\cite{2020.QUBIC.PAPER4}}.}

\newc{An isometric rendering of the cryostat is provided in the right-hand panel of figure~\ref{fig_intrument_system_overview}.} The cryostat weighs roughly 800~kg and is around 1.6~m high with a diameter of 1.4~m.

\new{The QUBIC-TD uses the same cryostat, cooling system, filters and general sub-system architecture as described above but with only 64~back-to-back horns and smaller mirrors to match the illumination of the $8\times8$~horn-array. It has a single 248~TES bolometer array operating at 150~GHz.}

\section{The horn-switch system}
\label{sec_horns_switches_system}

In the left panel of figure~\ref{fig_intrument_system_overview} we show a schematic of the QUBIC optical layout. The sky signal enters the cryostat propagating through a high-density polyethylene (HDPE) window. Then, a rotating half-wave plate modulates the polarization and a polarizing grid selects one of the two linear polarization components. An array of 400 back-to-back corrugated horns collects the radiation and re-images it onto a dual-mirror optical combiner that focuses the signal onto two orthogonal \new{focal planes populated by TES detectors that directly absorb the incoming radiation}. A dichroic \new{low-pass} filter placed between the optical combiner and the focal planes selects the two frequency bands, centered at 150 GHz and 220 GHz. The output of each detector contains interference fringes that are the so-called ``visibilities'' of the selected Fourier modes. \new{The detectors are unpolarized, so that the polarization of the detected radiation is determined by the combination of the polarizing grid with the rotating half-wave plate. The interested reader may find more information about the detectors system in \newd{Piat et al.} \newc{\cite{2020.QUBIC.PAPER4}}.}

\new{The right panel of figure~\ref{fig_intrument_system_overview} shows a sectional cut of the instrument solid model, while the bottom panel is a detailed view of a single element of the horn-switch array, consisting of two corrugated back-to-back horns connected by a circular, smooth \newb{waveguide} that can be shut by the switch blade (gray rectangle in the figure). The particular mechanical layout of the switch system (see figure~\ref{fig_single_switch_elements} in section~\ref{sec_switches_requirements_design}) did not allow us to place the blade at the center of the connecting \newb{waveguide}, but slightly closer to the input aperture.}



%
\section{Back-to-back feed horns}
\label{sec_back2back_prototype}

    \subsection{Feed horns requirements and design}
    \label{sec_feed_horns_requirements_design}

        \subsubsection{Requirements}
\label{sec_horn_requirements}

    \newd{The back-to-back horn array has two objectives: (1) the front (sky) horns define the field of view (FoV) of the instrument and (2) the back horns specify the edge taper illumination on the reflective optics and define the coherent inputs to the spatial beam combiner. A drawback of this implementation is that half of the power is lost in reflection to the sky by the polarizing grid. In table~\ref{tab_horn_requirements} we list the primary requirements of the back-to-back horn array with notes detailing their relevance. In our requirements we do not include the horn cross-polarization since the array is placed behind the polarization modulation/separation stage, so that the intrinsic horn cross-polarization is not a critical requirement.}

    \begin{table}[h!]
        \renewcommand{\arraystretch}{1.5}
        \caption{\label{tab_horn_requirements}Main requirements for the QUBIC back-to-back horns array}
        \begin{center}
            \begin{tabular}{m{4cm} m{2cm} m{8cm}}
                \hline
                Requirement& Value & Notes \\
                \hline
                \hline
                \new{Center-to-center} distance\dotfill & 14\,mm & \new{The smallest achievable, for the optimal sampling} of the angular power spectrum.\\
                Aperture\dotfill & 12\,mm & Driven by the FoV \new{and by the illumination of the optical combiner}.\\
                Return loss\dotfill & $<-25$\,dB & Over the 130--240\,GHz bandwidth. \new{This is the target RL value for the fundamental mode only, used to optimize the horn design.}\\
                Insertion loss\dotfill & $<0.1$\,dB & To ensure a negligible contribution to optical losses, the \new{overall} transmission being \new{higher than 60\%}.\\
                Mass\dotfill & 30\,g/horn & Must be suspended on the top of the optical combiner \new{and operated in a cryogenic environment.}\\
                \hline
            \end{tabular}
        \end{center}
    \end{table}

\subsubsection{Mechanical and electromagnetic design}
\label{sec_design}

    \newd{We built identical front and back horns with a design based on  an  initial  study  by  Maffei  et  al.  \cite{maffei2016} that  explored  the performance of waveguide feed horns with a corrugated ﬂare proﬁle inspired by a Winston concentrator geometry.}  {This profile was modified and further developed for the project Clover \newc{\cite{Clover2008}} and also adopted for other instruments such as QUIJOTE \newc{\cite{quijote2010}}, and BINGO \newc{\cite{BINGO2020}}.} 
    
\begin{figure}[h!]
        \begin{center}
            \includegraphics[width=14cm]{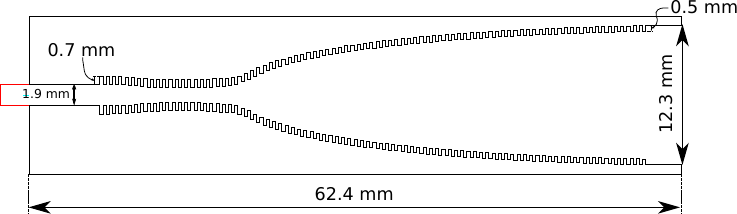}
        \end{center}
        \caption{\label{fig_corrugation_profile}The corrugations profile of the QUBIC horns.}
    \end{figure}

    \new{The profile optimized for QUBIC (see figure~\ref{fig_corrugation_profile}) has 0.5\,mm deep corrugations to reduce the cross-polarization and return loss at 150\,GHz. The horn \newb{waveguide} has a corrugation depth of 0.7\,mm and a minimum radius of 0.684\,mm so that we select the HE$_{11}$ for the 150\,GHz band with a cut-off frequency of about 130\,GHz. The guide design also allows the propagation of four modes up to 240\,GHz band, with the drawback of a larger cross-polarization at this frequency.}
    
    \new{We also added a transition between the Winston and the \newb{waveguide} sections of the horn which have a different corrugation depth, and a re-expansion that matches the corrugated \newb{waveguide} to the smooth \newb{waveguide} of the switch block.}
    
    \new{As already noticed, although QUBIC is largely insensitive to any cross-polarization injected after the polarizing grid, we chose to implement a corrugated horn design that is generally used in single-mode optical systems with stringent requirements in cross-polarization and side-lobes. To understand our choice it is necessary to place it in the framework of the historical development of QUBIC and of the main changes in its design occurred over the years.}

    \new{The choice of corrugated horns for QUBIC dates back to early phases of the project, described, e.g., in \newd{QUBIC collaboration et al.} \newc{\cite{2011APh....34..705Q}}, when QUBIC was designed as an ensemble of single-band, dual-polarization instruments, with one dedicated cryostat per frequency band. At that time the polarizing grid  was placed at 45$^\circ$ after the optical combiner to split the polarization states onto the two focal planes. In this design the feed-horn cross-polarization was an important issue and, therefore, we considered corrugated horns the best choice.}
    
    \new{After the BICEP2 claim in 2014 \newc{\cite{bicep2-2014}}, the importance to have a frequency channel dedicated to dust monitoring became clear and, consequently, the collaboration decided to modify the design to allow the detection of both 150\,GHz and 220\,GHz bands in the same module. This led to the current design, with the polarizer grid immediately below the rotating half-wave plate and a dichroic filter that splits the two frequency bands.}
    
    \new{Meanwhile, a few prototypes of the corrugated horns had already been produced for a production chain for this design to be in place. For this reason we decided to modify just the most critical part of the horn, the mode converter, to allow the propagation of both the 150\,GHz and the 220\,GHz band, leading to the current geometry in which the horns are single-moded at 150\,GHz and multi-moded at 220\,GHz.}

\subsubsection{Electromagnetic simulations technique}
\label{sec_simulation_technique}

\begin{figure}[h!]
        \begin{center}
            \includegraphics[width=10cm]{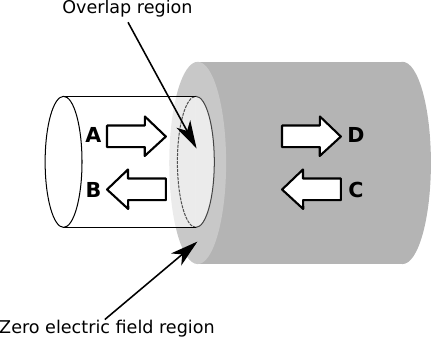}
        \end{center}
        \caption{\label{fig_mode_matching}Schematic of the mode-matching model implemented in the electromagnetic simulations.}
    \end{figure}

    We have simulated the fields propagating from the \newb{waveguide} to the aperture of the QUBIC corrugated horn using the electromagnetic mode-matching technique \newc{\cite{james1981,clarricoats}}, depicted schematically in figure~\ref{fig_mode_matching}. This technique regards the corrugated structure as a sequence of smooth walled cylindrical \newb{waveguide} sections, each of which can support a set of propagating \new{transverse electric (TE) and transverse magnetic (TM)} modes. At each corrugation the sudden change in the radius results in a scattering of power into backward propagating reflected modes in the left-hand side guide segment and forward propagating transmitted modes in the right-hand segment. 

    The power coupling between modes is given by the overlap integral $\int e_{n,l}\, h_{m,r}\, dA$, where $e_{n,l}$ is the transverse electric field of mode $n$ on the left-hand side of the junction, $h_{m,r}$ is the magnetic field of mode $m$ on the right-hand side of the junction and $dA$ is a surface element on the transverse plane. The modes are then propagated through the length of \newb{waveguide} section to the next scattering junction where the overlap integral between the modal components is computed again. 

    If $\vec A$ and $\vec C$ are column vectors of the mode coefficients of the fields incident from the left and the right, and $\vec B$ and $\vec D$ are the mode coefficients of the resulting reflected fields, then their relationship is described using a scattering matrix, $\mathbf{S}$:
    \begin{equation}
        \begin{bmatrix}
        \vec B \\
        \vec D
        \end{bmatrix}
        = \mathbf{S}\cdot
        \begin{bmatrix}
        \vec A \\
        \vec C
        \end{bmatrix}
        =
        \begin{bmatrix}
        \mathbf{S_{1,1}} & \mathbf{S_{1,2}} \\
        \mathbf{S_{2,1}} & \mathbf{S_{2,2}}
        \end{bmatrix}
        \cdot
        \begin{bmatrix}
        \vec A \\
        \vec C
        \end{bmatrix}
    \end{equation}
    whose elements are calculated using overlap integrals as described in \newd{Olver et al.} \newc{\cite{olver}}. The columns of the scattering matrix describe the amplitude of each output mode generated by a unit-amplitude input mode. The scattering matrix for the horn as a whole, is computed by cascading the matrices for each uniform section and junction. We assume no scattering at the horn aperture, \new{since it is electrically large (6$\lambda$)}, so $\vec C=0$.

    The field at the aperture of the corrugated horn is then determined from $\vec D = \mathbf{S_{2,1}}\cdot \vec A$, where $\mathbf{S_{2,1}}$ is the sub-matrix that deals with the forward-propagating modes, and the reflected field is determined from $\vec B = \mathbf{S_{1,1}}\cdot \vec A$. The transmitted and reflected power are found by multiplying the complex elements of the relevant column vector by their complex conjugate and summing them.
    
    In our analysis we used 60 \newb{waveguide} modes (30 TE and 30 TM), \new{which was more than sufficient for the accuracy levels shown in this paper.} The TE and TM modes with power have a coherent phase relationship and in this case correspond to the single hybrid $\mathrm{HE}_{1,1}$ mode.
    
    In the 220 GHz band more than one column of the scattering matrices is non-zero and these represent possible independent modes of power transmission.  We excite all modes equally at the input, $\vec A = [1, 1, 1, \ldots]^T$, and add the individual output fields incoherently (a detailed description can be found in \newd{Murphy et al.} \newc{\cite{murphy1991,murphy2001}}). The reflected power is calculated as a percentage of the power that could be transmitted by the number of propagating modes. Spikes can be observed at cut-off frequencies of higher order modes as these states transition from evanescent to propagating modes. \new{These predicted features are also likely to depend on the exact details of the manufactured horns.}

\subsubsection{Simulations results}
\label{sec_simulation_results}

\begin{figure}[h!]
        \begin{center}
            \begin{tabular}{c c c}
            \includegraphics[width=7cm]{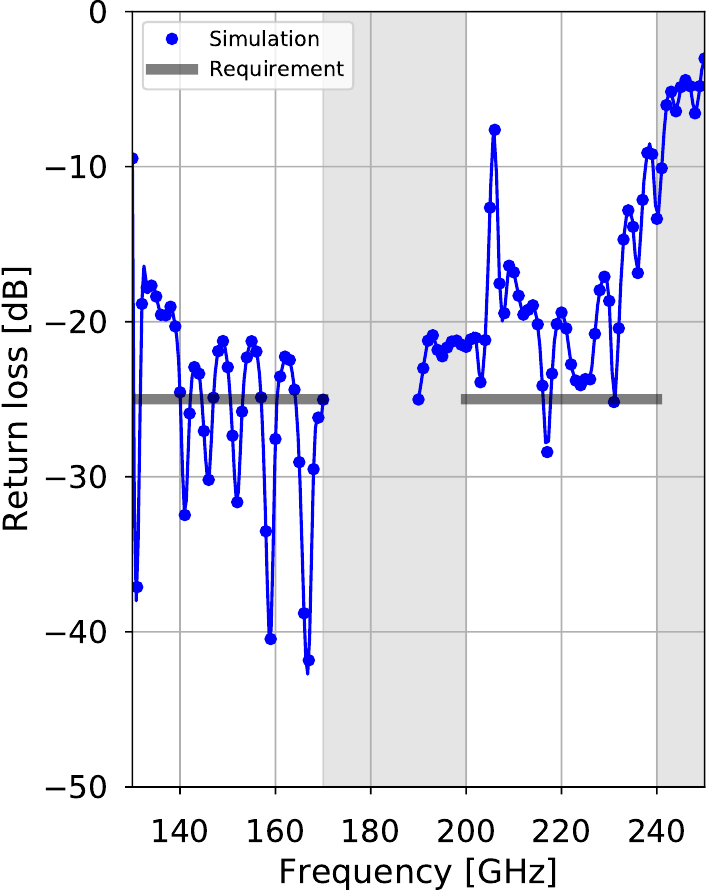} &
            \,\,\,&
            \includegraphics[width=7cm]{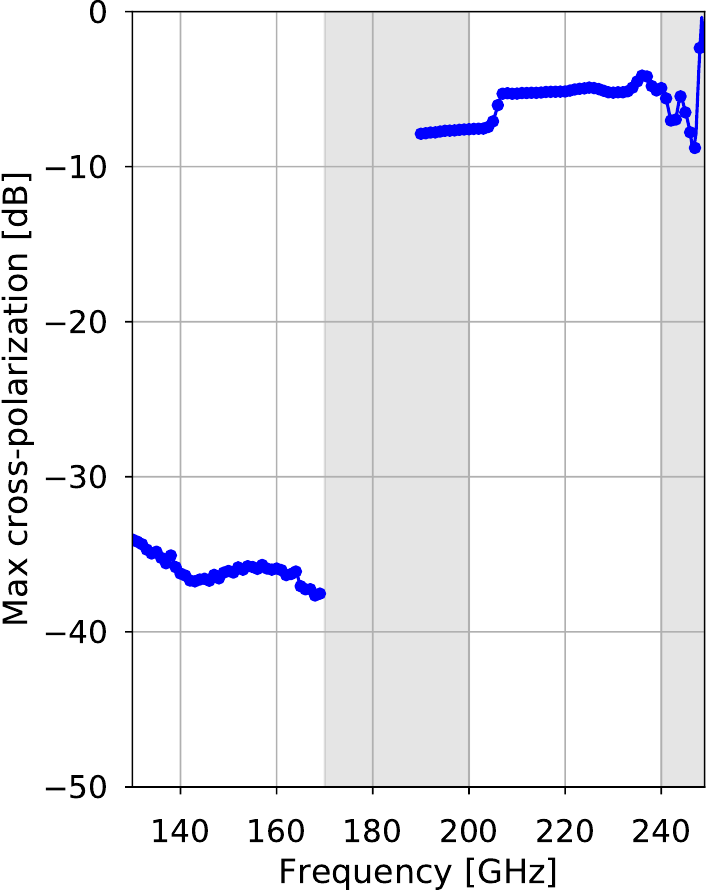}
            \end{tabular}
        \end{center}
        \caption{\label{fig_horn_simulated_return_loss_xpol}Simulated return loss (left) and maximum cross-polarization (right) in the two QUBIC frequency bands.}
    \end{figure}

    In figure~\ref{fig_horn_simulated_return_loss_xpol} we show the return loss (left panel) and maximum value of the cross-polar beam (right panel) in the two QUBIC bands. We can see at a glance that the performance in the 150\,GHz band is superior compared to the 220\,GHz band. In fact the design was initially tailored in the D-band and subsequently modified to accept also the higher band that could not be optimized in terms of performance like the lower frequency range.

    The return loss at 150\,GHz is, on average, around $-$25\,dB, while in the higher frequency band it is compatible with $-$20\,dB up to 230\,GHz, and degrades to $\sim -10$\,dB on the right hand side of the frequency interval. We assessed the potential impact of the poor return loss in the highest part of the 220\,GHz band: a degradation of the return loss will induce a reduction of the horn transmission and therefore an overall decrease of the sensitivity. \new{With a pessimistic $-$10\,dB return loss over the whole 220\,GHz, we estimate a degradation in the sensitivity of about 5\%. \newd{We accepted this violation of the speciﬁed requirement because of the challenges in simultaneously optimizing the corrugated horns for both 150\,GHz and 220\,GHz bands without a signiﬁcant increase in fabrication complexity.}}

    The cross-polarization is very good ($\sim -35$\,dB) at 150\,GHz, while it is around $-$5\,dB at 220\,GHz. This is \new{in agreement} with the design: at 150\,GHz we have a single-mode corrugated horn, for which we expect excellent polarization purity, while at 220\,GHz we have propagation of higher modes that do not preserve the polarization state. But this is not a problem for QUBIC, as already mentioned, because the polarization is selected before the radiation enters the horns. For this reason we show here the expected cross-polarization performance but we will not discuss it further in the rest of the paper.

\begin{figure}[h!]
        \begin{center}
            \includegraphics[width=14cm]{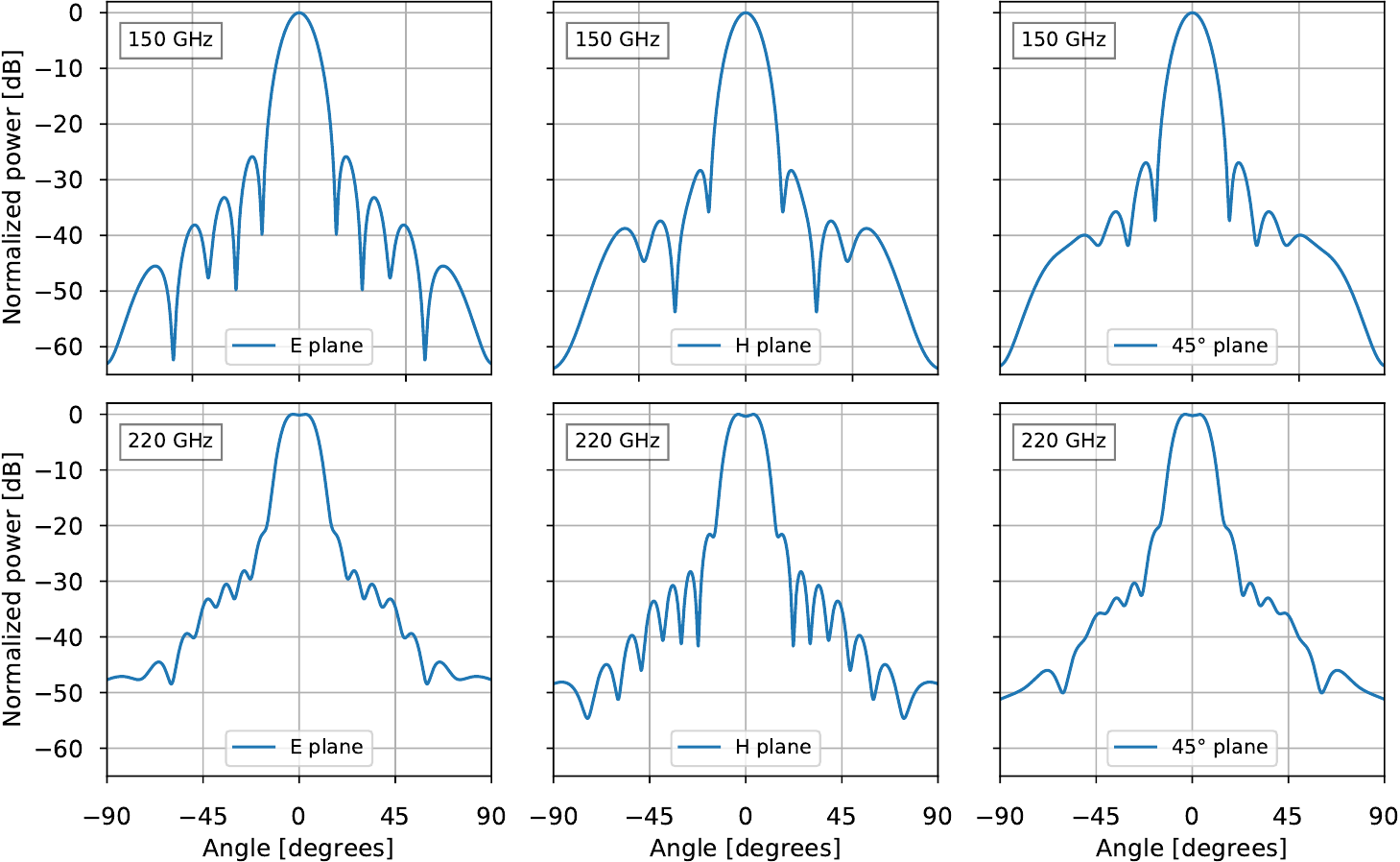}
        \end{center}
        \caption{\label{fig_horn_simulated_pattern}Co-polar simulated beam patterns (E, H, and 45$^\circ$ planes) at 150\,GHz (top row) and 220\,GHz (bottom row).}
    \end{figure}

    In figure~\ref{fig_horn_simulated_pattern} we show the simulated \newb{normalized} beam patterns at 150 and 220\,GHz for the three main co-polar planes (E-plane, H-plane and 45$^\circ$ plane). \newb{The maximum directivity in the two frequency bands is 23.8\,dBi at 150\,GHz and 22.4\,dBi at 220\,GHz.} At 150\,GHz we can appreciate the typical Gaussian profile of single-mode corrugated horns, while at 220\,GHz the main beam shape is a flat-top resulting from multi-mode propagation. The side-lobes are low, less than $\sim-$30\,dB at angles larger than $\sim 30^\circ$.

    \subsection{The technological demonstrator feed horn array}
    \label{sec_td_horns}
    
        \new{We have chosen to develop the horn arrays in platelets because this technique represents the best option to realize large arrays of corrugated horns at our frequencies with the required performance and a sustainable cost, when compared with other options like electroforming, electrical discharge machining or even direct machining.}

\new{This technique, introduced by Haas in 1993 \cite{haas1993,haas1995}, requires drilling circular holes into metal plates that are subsequently stacked and clamped or bonded. For QUBIC we have used two methods to drill the holes: chemical etching of 0.3\,mm aluminum plates and computer controlled milling of top and bottom plates (3\,mm and 6\,mm thick, respectively). Then we mechanically clamped the plates with screws \newb{(applying a torque of 0.8\,N$\cdot$m)} and alignment pins. We have shown in other works \newc{\cite{deltorto2011,mandelli2021}} that mechanical clamping is a valid alternative to bonding, does not impact negatively the performance and can be used also when the antennas operate in cryogenic conditions. \newb{In particular in \cite{deltorto2011} Del Torto et al. showed that a clamped platelet array of W-band feedhorns maintained its return loss performance after cooling to 4\,K.}}

\begin{figure}[h!]
    \begin{center}
        \includegraphics[width=15cm]{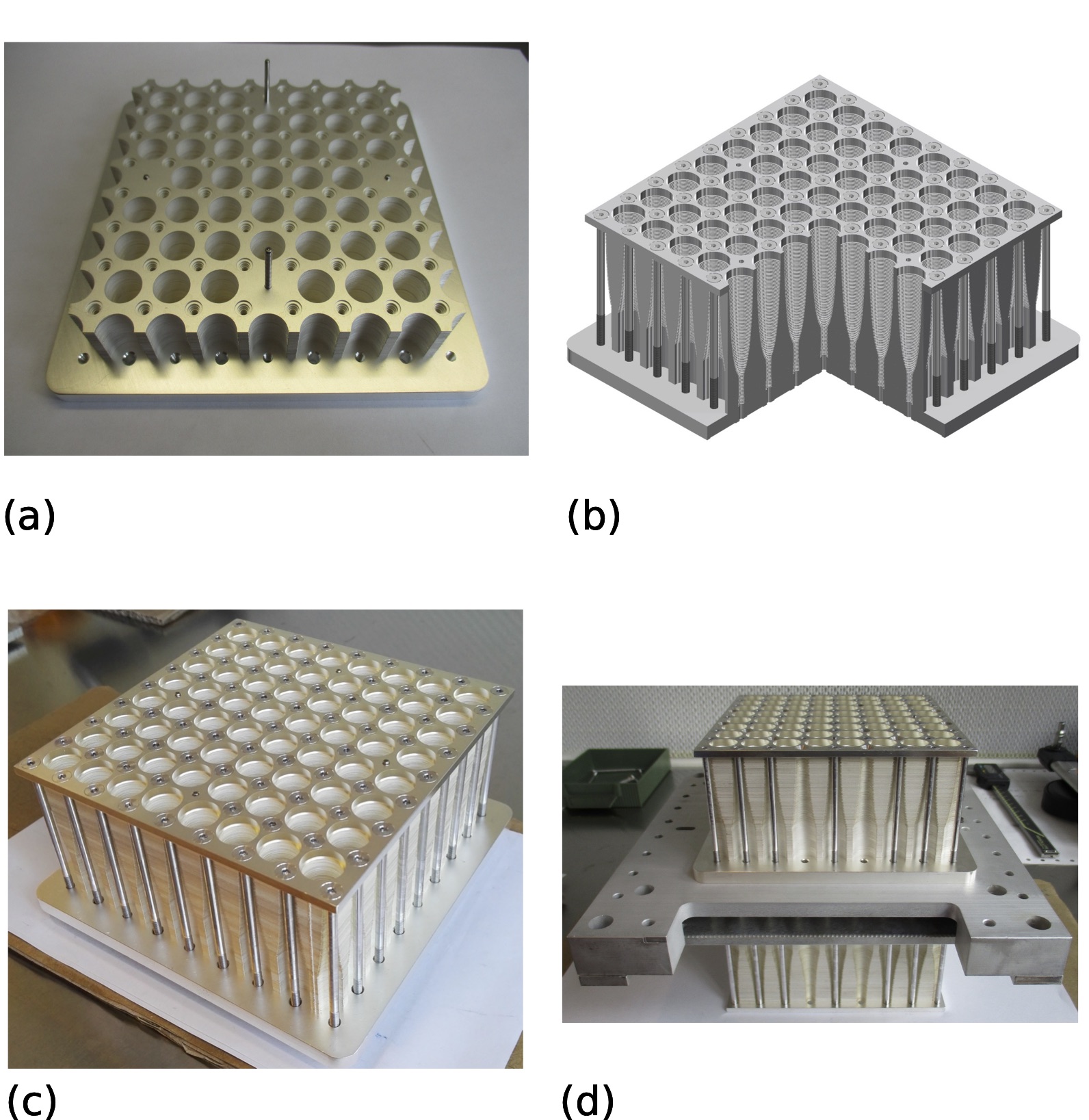}
    \end{center}
    \caption{\label{fig_qubic_horns_mechanical_development}The QUBIC TD feed-horns array: (a) chemically etched aluminum platelets during the stacking process; (b) 3-D model of the antenna array; (c) one of the two antenna modules after integration; (d) the complete integrated feed-horn-switch system.}
\end{figure}

Each of the two 64-horns arrays is composed of 175 plates enclosed between the external flanges (panels (a), (b), and (c) of figure~\ref{fig_qubic_horns_mechanical_development}). We silver-plated each plate and both flanges to improve electrical conductivity along the horns profile. \newe{The thickness of the silver plating (quoted by the manufacturer) is $3.5\,\mu\mathrm{m}\,\pm\,0.3\,\mu\mathrm{m}$ and the r.m.s. surface  roughness measured in our workshop on one of the plates is $R_\mathrm{q}=0.3\,\mu\mathrm{m}$.} \new{Indeed, on previous prototypes we verified the effect of silver plating by measuring the room-temperature insertion loss of platelet horns with and without silver plating. We observed that the insertion loss decreased from about 0.45\,dB to less than 0.1\,dB when the aluminum plates were silver coated.}

Each metal plate has an overall size of 112\,mm$\times$112\,mm, with a thickness of 0.3\,mm. The plates also include four 3\,mm diameter holes for alignment pins and 77 holes for the ERGAL (7075 aluminum alloy) M3 screws that pack the array between the two flanges. 

The top flange has an overall size of 120\,mm$\times$120\,mm with a thickness of 3\,mm and it contains the 64 horn apertures with a diameter of 12.33\,mm. It also contains countersink holes for the tightening screws so that the screw heads lie flush with the array planar surface. The bottom flange has an overall size of 133\,mm$\times$133\,mm with a thickness of 6\,mm and it contains circular \newb{waveguide} segments with a diameter of 1.91\,mm for each of the 64 horns. 

The bottom flange couples to the switch array and is machined in a shape similar to that of the UG-387 waveguide flange (MIL-DTL-3922/67C) to obtain anti-cocking\footnote{\newc{The ``cocking'' eﬀect refers to the presence of a non-parallel ﬂange mate or gap at the interface, which can lead to signal loss and mechanical damage.  The geometry of the ``anti-cocking'' interface is designed to minimize this possibility during installation of the waveguide ﬂange fasteners \cite{Kerr2009}.}} interfaces between each \newb{waveguide} pair.
It also includes holes for the module tightening screws. Half of the screws run from the top to the bottom flange and hold each of the two arrays together (therefore allowing us to handle and test the arrays separately), the remaining group extends further to hold the horn and switch modules together (see panel (d) of figure~\ref{fig_qubic_horns_mechanical_development}).

%
    \subsection{Mechanical measurements and achieved tolerance}
    \label{sec_second_prototype_metrology}

          \subsubsection{Experimental procedures}
  \label{sec_metrology}
  
	We tested the mechanical tolerance according to two different procedures. First we visually inspected the inner profile of a 4-elements horn array manufactured in brass that was cut to allow us to magnify the shape of the antenna teeth and grooves. Then we used a metrological machine (Werth ScopeCheck 200) to measure the position, diameter and deviation from circularity of each hole in the platelets of the final array.
    
\begin{figure}[h!]
      \begin{center}
        \includegraphics[width=7cm]{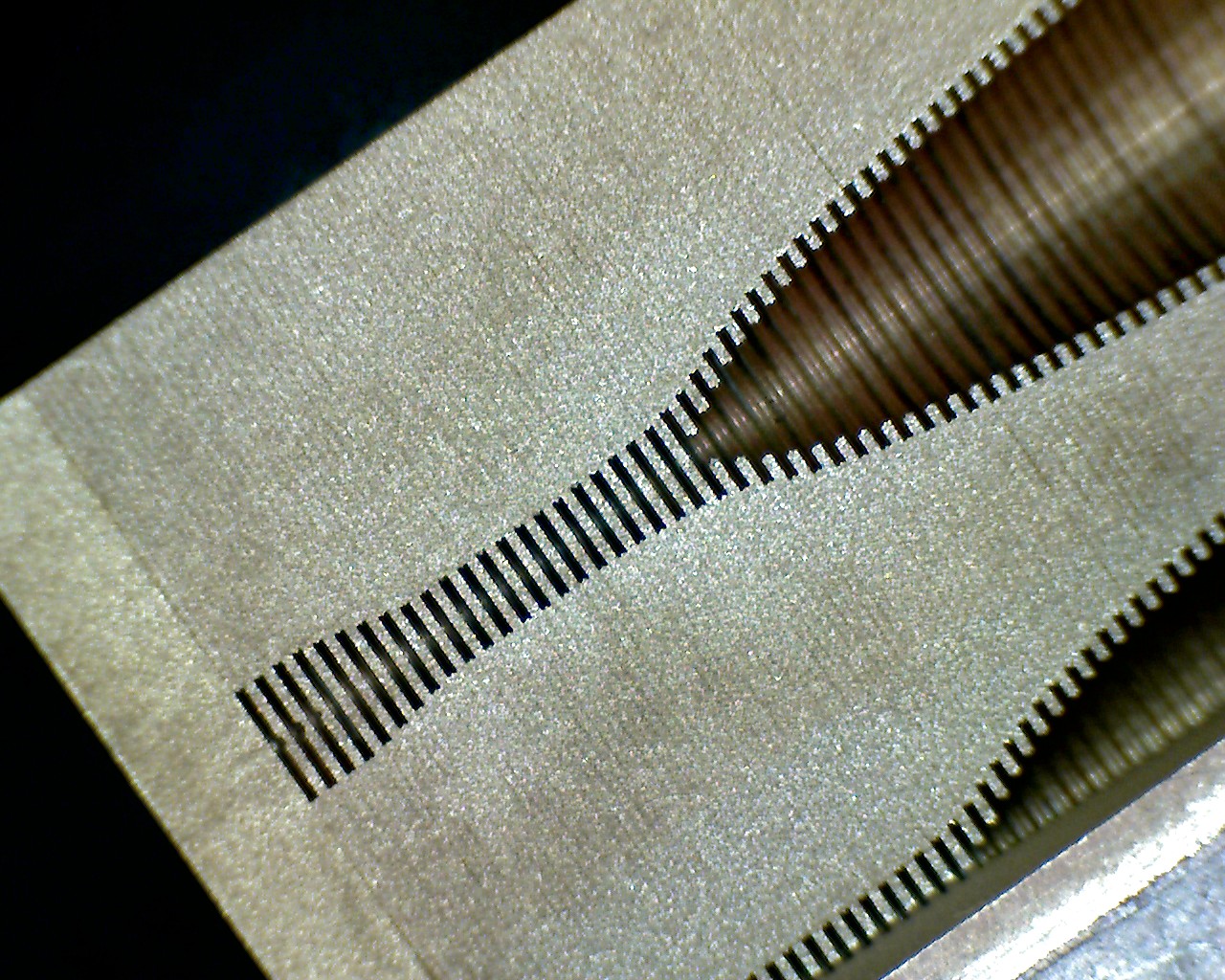}
        \includegraphics[width=7cm]{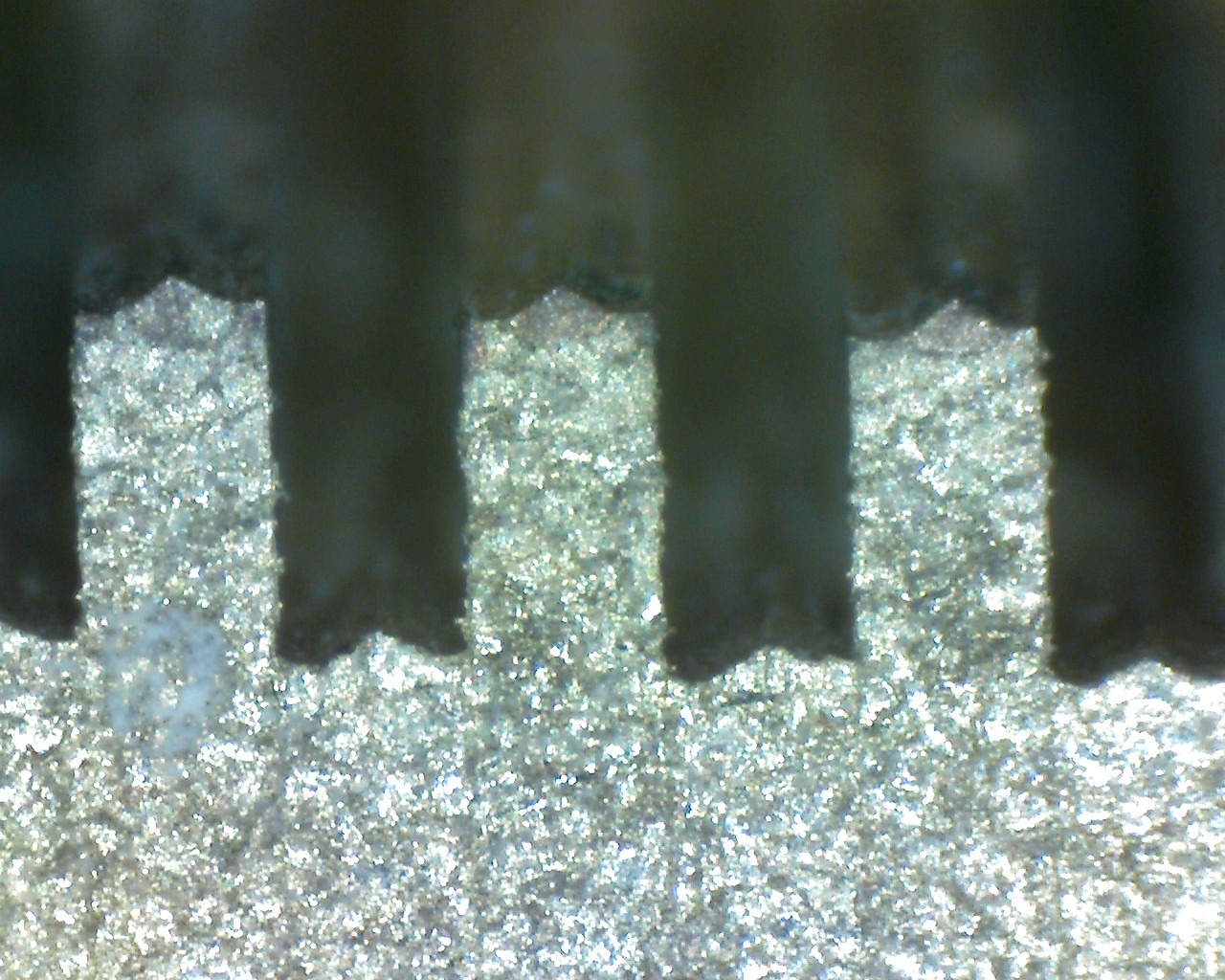}
      \end{center}      
      \caption{\label{fig_details_of_apices}\textit{Left.} A section of the first feed-horn prototype. \textit{Right.} Detail of the cusps on teeth and grooves resulting from non uniform chemical erosion.}
    \end{figure}

	\paragraph{Visual inspection.} The left panel in figure~\ref{fig_details_of_apices} shows a section of the brass prototype. The enlargement in the right panel highlights the presence of cusps ($\lesssim 0.06$\,mm high) on the profile of all the corrugations that are the effect of a non uniform erosion of the metal during the etching process. 
    
    This non uniformity is a limitation which is inherent in the chemical etching process, so that we can expect that the antennas produced with this method present imperfections in their corrugated profile. In section~\ref{sec_impact_mechanical_imperfections} we discuss the impact of these defects on the feed-horn performance for QUBIC.

	\paragraph{Metrology.} \newc{We carried out metrological measurements using the Werth ScopeCheck 200 instrument,} that performs precision measurements using either an optical or a tactile device. In our setup we used the optical sensor, which can be moved in three dimensions over a glass work plane where we laid our platelets. In section~\ref{sec_metrology_results} we discuss the results of the metrological measurements.

%
  \subsubsection{Impact of mechanical imperfections on electromagnetic performance}
  \label{sec_impact_mechanical_imperfections}
    
\begin{figure}[h!]
    	\begin{center}
        	\includegraphics[width=10cm]{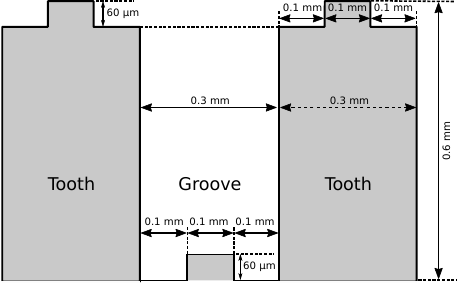}
        \end{center}
         \caption{\label{fig_cusp_model}Sketch \newc{(drawn to scale)} of the model used to simulate the imperfections in the feed-horn profile.}
    \end{figure}

    We assessed the impact of the imperfections in the feed-horn profile caused by the etching process by computing the return loss and the co-polar radiation patterns on the $E$ and $H$ planes considering two cases: (i) the nominal profile, and (ii) a profile modified inserting a step-like defect on teeth and grooves of all the corrugations (see figure~\ref{fig_cusp_model}). 
    
    \new{We notice that the step-like defect in figure~\ref{fig_cusp_model} is not fully representative of the cusp-like structure shown in figure~\ref{fig_details_of_apices}. On the other hand we had to use the step-like approximation because the mode-matching code we used to simulate the multi-moded beams does not allow the simulation of cusp-like structures. \newb{We validated the use of this simpliﬁcation by comparing single-mode beam simulations carried out with two codes:  our mode-matching code for the step-like model and the SRSR software \newc{\cite{SRSR}} for the cusp-like structure.} }
    
    \paragraph{Return loss.}
    \label{sec_return_loss_cusps}
    
\begin{figure}[h!]
      \begin{center}
        \includegraphics[width=14cm]{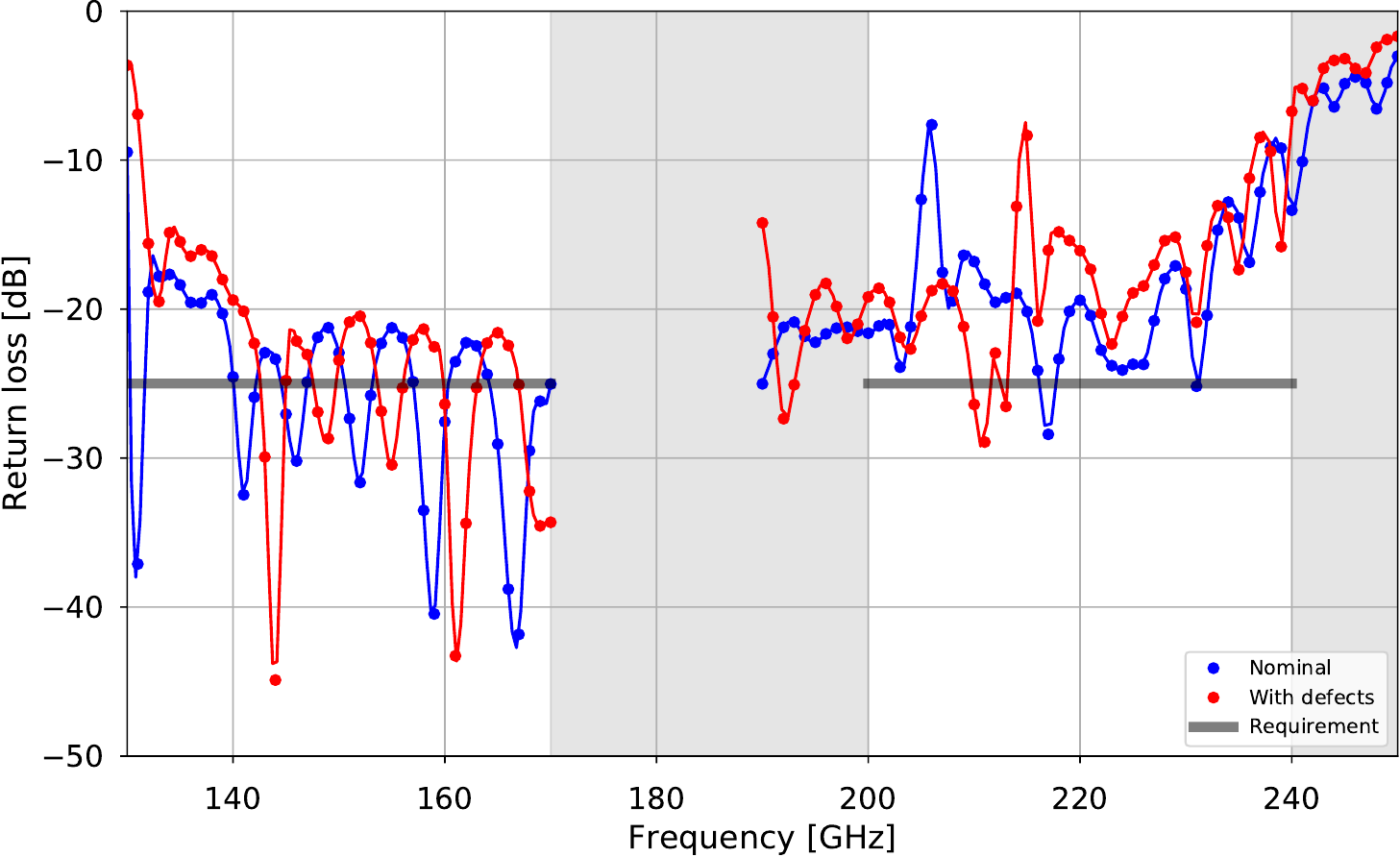}
      \end{center}
      \caption{\label{fig_rl_cusps}Impact of defects on the feed-horn return loss.}
    \end{figure}

    Figure~\ref{fig_rl_cusps} shows the effects of the defects on the return loss\new{, and we see that the narrower effective size of the etched holes shifts the resonances in frequency}. We see that they do not change significantly the overall level, but shifts some of the resonances in frequency. In general, however, we can consider the impact on the return loss negligible. 
      
    \paragraph{Radiation pattern at the center frequency.}
    \label{sec_pattern_simulation}      
    
\begin{figure}[h!]
      \begin{center}
        \includegraphics[width=14cm]{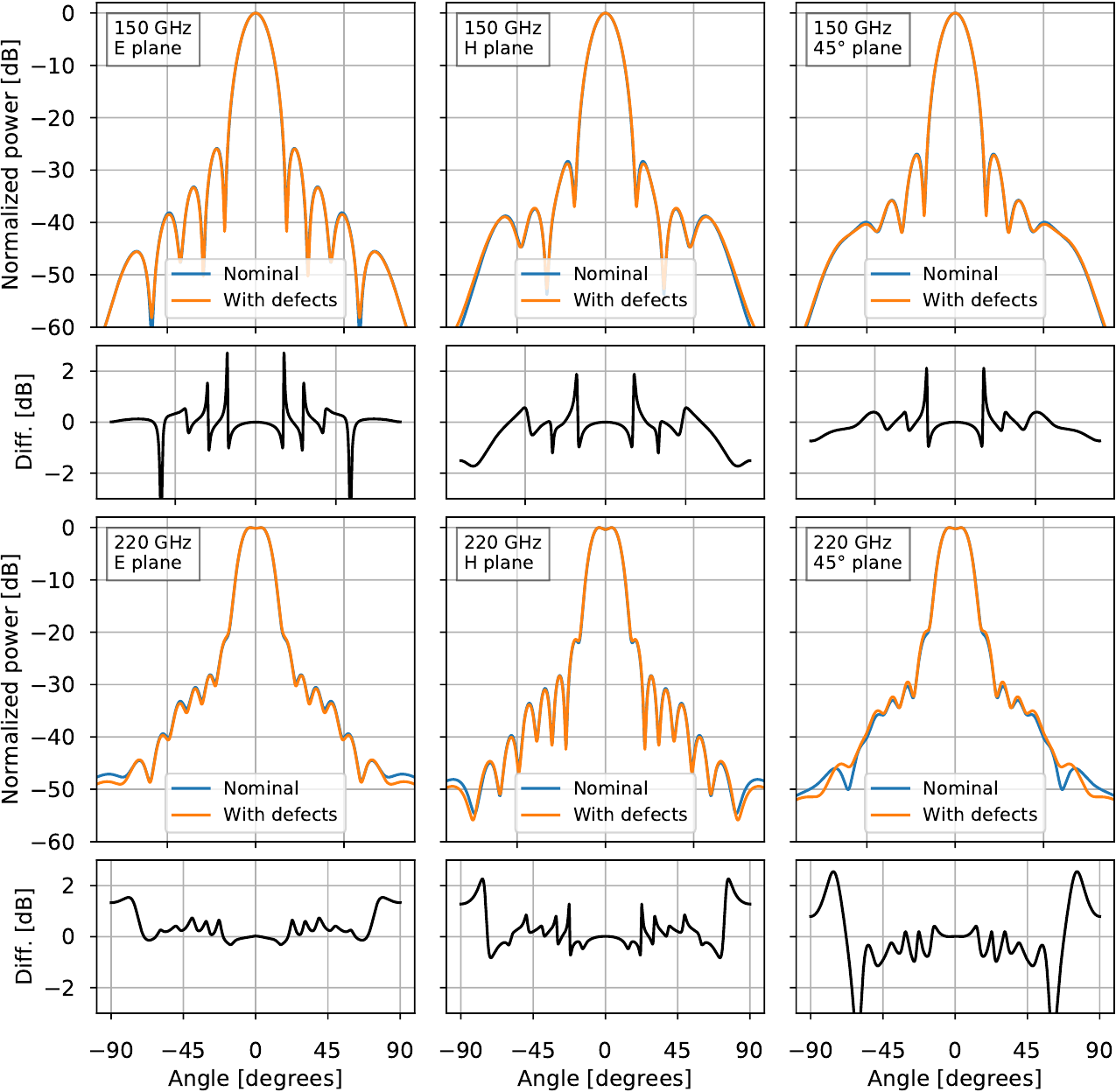}
      \end{center}     
      \caption{\label{fig_beam_cusps}Impact of defects on the co-polar radiation patterns. The black line in the bottom panel shows the difference between the two beam patterns. The top plots refer to 150\,GHz simulations, the bottom plots refer to 220\,GHz simulations.}
    \end{figure}

      Figure~\ref{fig_beam_cusps} shows the simulated radiation patterns (E-plane, H-plane, and 45$^\circ$ plane) at 150\,GHz (top rows) and 220\,GHz (bottom rows) for the two cases studied. The bottom plot in each figure shows the difference in dB of the beam patterns for the two cases. In the main beam region ($-15^\circ < \theta < 15^\circ$) the difference is less than 0.05\,dB, and over all the $-90^\circ \leq \theta \leq 90^\circ$ range the difference is within $\pm$2\,dB.
 
  \subsubsection{Results of metrological measurements}
  \label{sec_metrology_results}
  
    We have carried out metrological measurements of both feed-horn arrays and compared the manufacturing precision with the maximum achievable tolerance of the chemical etching process, which is $\pm$0.05\,mm. In this section we will refer to the two arrays as array-1 and array-2.

\begin{figure}[h!]
      \begin{center}
        \includegraphics[width=10cm]{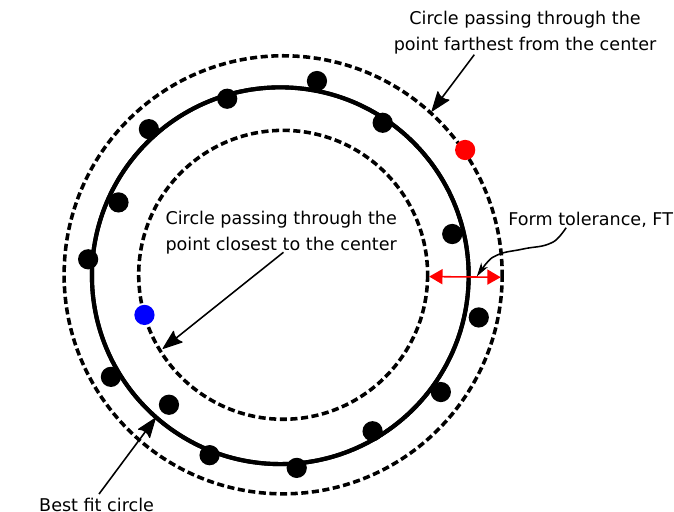}
      \end{center}     
      \caption{\label{fig_form_tolerance}Definition of the form tolerance (FT) parameter.}
    \end{figure}

    We measured the holes of each antenna and alignment pin for all the aluminum plates, compared the measured positions and diameters with their nominal values and calculated the form tolerance (FT) of each hole (see the sketch in figure~\ref{fig_form_tolerance} for a definition of this parameter). This rich set of measurements allowed us to obtain the actual mechanical profiles of all the feeds in the array that we used to simulate their actual electromagnetic behavior. We then compared this family of simulations with the electromagnetic parameters measured in the laboratory, as explained at the beginning of section~\ref{sec_electromagnetic_measurements}.
    
\begin{figure}[h!]
        \begin{center}
            \includegraphics[width=7cm]{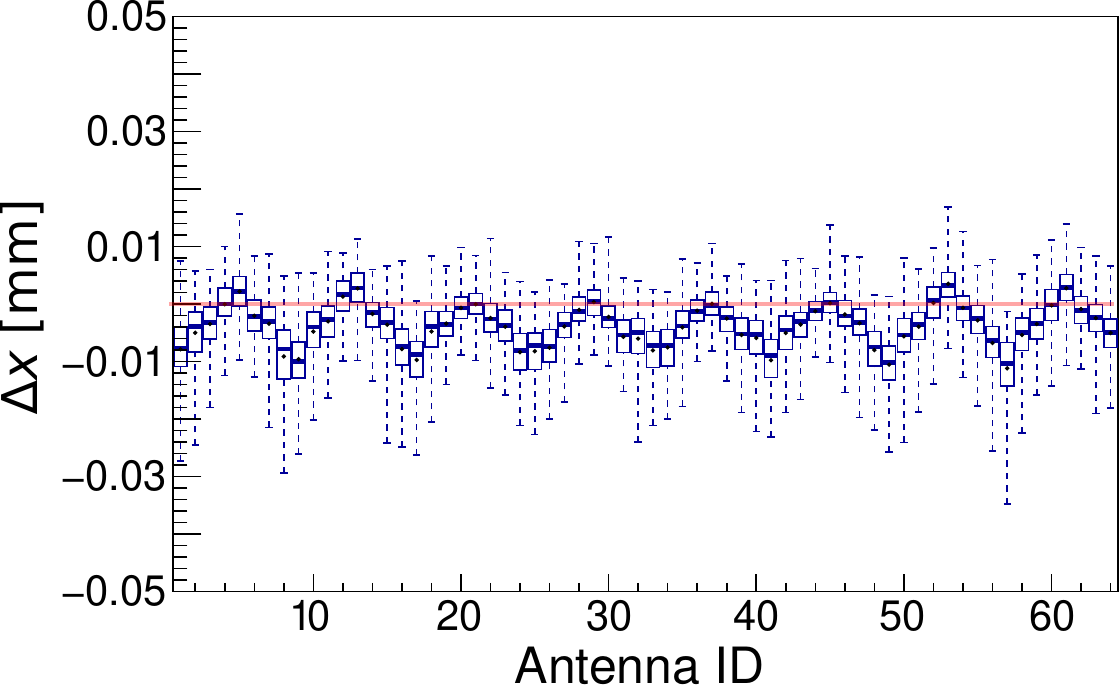}
            \includegraphics[width=7cm]{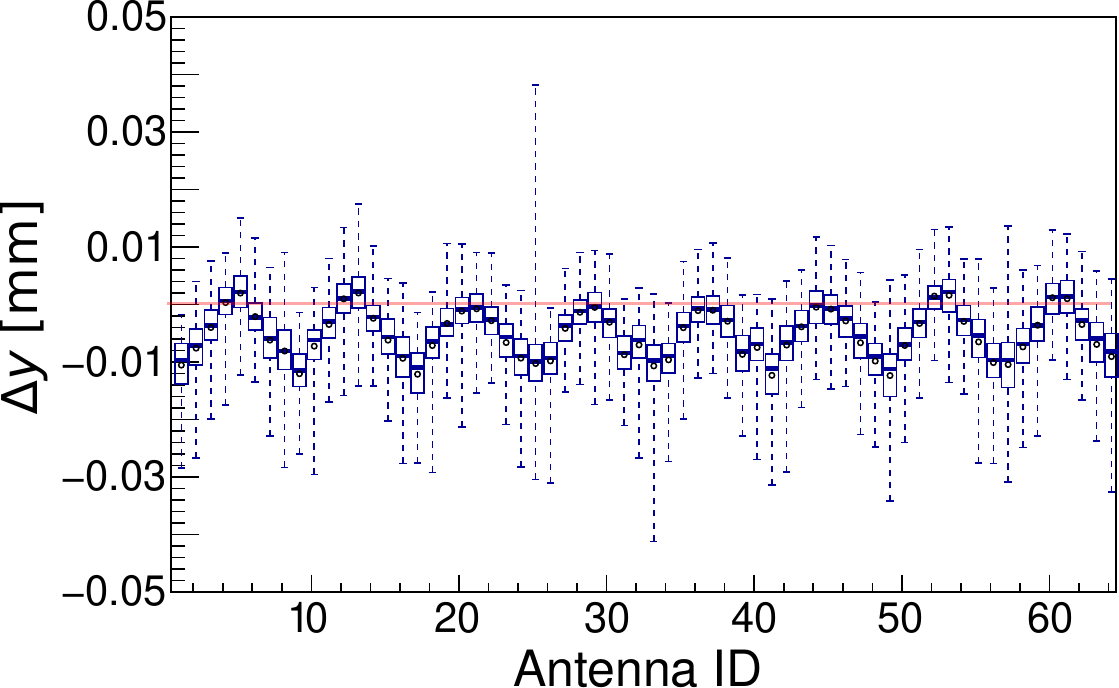}
            \includegraphics[width=7cm]{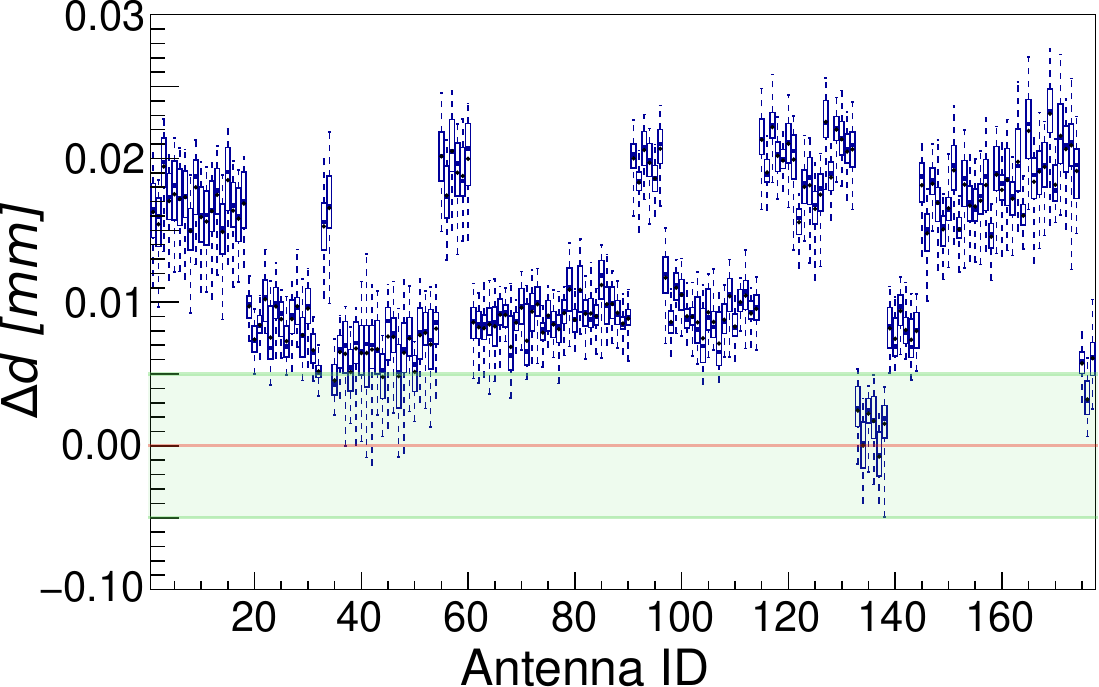}
            \includegraphics[width=7cm]{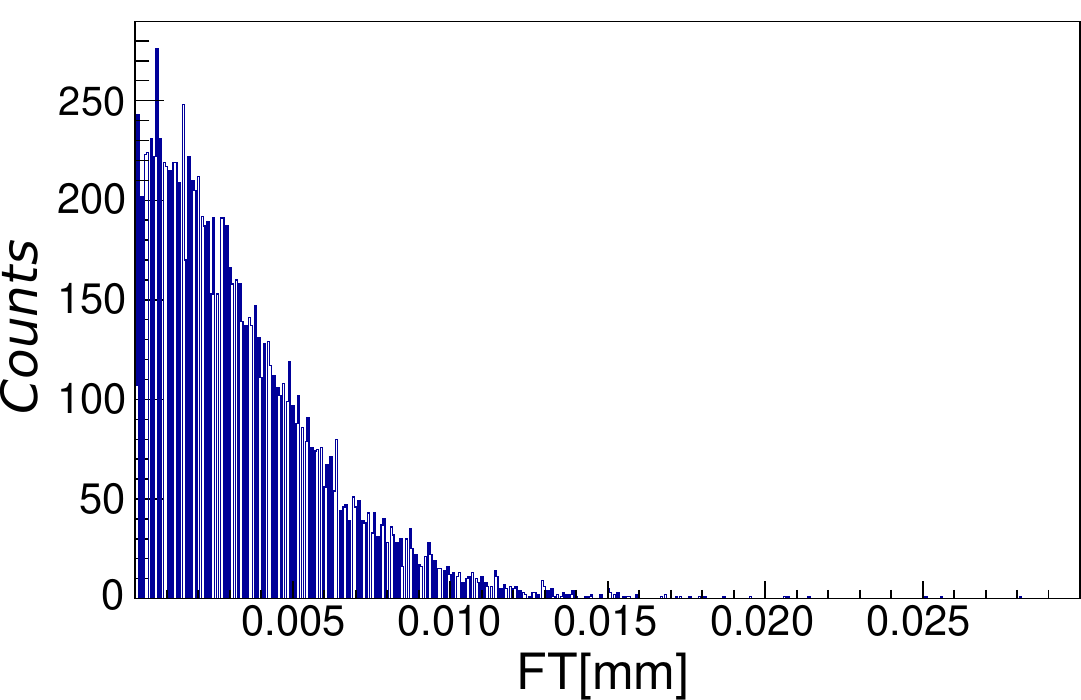}
        \end{center}
        \caption{\label{fig_metrology_results}Results of the metrological measurements of the antenna holes of array 1. \new{We do not report the results for the second array because they tell essentially the same story.} \textit{Top-left} and \textit{top-right}: deviation of measured $x$ and $y$ center coordinates from the nominal value. \new{Here for each antenna (horizontal axis) we display the spread of the deviations, $\Delta x$ and $\Delta y$, of the holes center coordinates. The box represents the 1-$\sigma$ spread, the dashed line the peak-to-peak. The red line corresponds to a null deviation.} \textit{Bottom-left:} deviation of measured diameters from the nominal value. \new{Here for each plate (horizontal axis) we show the spread of the deviations, $\Delta d$, of the holes diameters. Again, the box represents the 1-$\sigma$ spread, the dashed line the peak-to-peak. The red line corresponds to a null deviation, while the green area highlights the expected manufacturing tolerance.} \textit{Bottom-right}: measured form tolerances \new{for all the holes in the antenna array platelets. See figure~\ref{fig_form_tolerance} for the definition of form tolerance}.}
    \end{figure}

    The box-plots of figure~\ref{fig_metrology_results} show the deviation of the measured Cartesian center coordinates of the antenna holes from their nominal value ($\Delta x$, top-left, and $\Delta y$, top-right), the deviation between the measured and nominal hole diameters (bottom-left), and the corresponding distribution of FT values (bottom-right). The red line corresponds to a null deviation, while the green area highlights the expected manufacturing tolerance. The measurements are relative to the array-1.

    As one can see, all the antenna positions comply with the manufacturing tolerance, while more than $90\%$ of the antenna diameters are out of specification, generally larger than expected and distributed around two peaks: $\Delta d_1$ = 0.07\,mm and $\Delta d_2$ = 0.15\,mm, with a maximum deviation of 0.25\,mm. The measured FT show no significant deviation from circularity.

    We obtained similar results for the alignment pins of array-1 and for both antenna holes and pins of array-2, but they are not reported here for simplicity. We measured also the top and bottom plates of the arrays and they are in compliance with the milling precision tolerance of 0.03\,mm.
    
    The out-of-spec was due to a loose control of the chemical etching time. Indeed, one can see that the measurements are grouped in blocks of plates and the average deviations from the nominal diameters follow a bi-modal distribution. Discussing with the company that performed the etching we understood that the plates were treated in batches \new{and that the etching time was not carefully controlled among the various batches for reasons independent of QUBIC}. These problems were solved in the production of the horns for the FI (see section~\ref{sec_fi_horns}) by strictly controlling the etching time.
    
    In section~\ref{sec_electromagnetic_measurements} we discuss the effect of this out-of-spec on the TD horns electromagnetic performance, where we compare the measured return loss and beam patterns with simulations run with the nominal and measured antenna profiles.

    The box-plots in figure~\ref{fig_metrology_results} also highlight an oscillatory, almost sinusoidal pattern in the measurements of the holes centers coordinates as a function of the antenna number. This is likely a systematic effect in our measurement. In fact, this behavior correlates with the row-by-row scanning of the antenna holes in the square antenna array. \newc{We found a similar oscillatory pattern also in the diameter measurements, with an amplitude much smaller than the spread in the deviation of the diameters from their nominal value. Because we could not clearly identify this effect, neither in the measurement strategy nor in the measurement machine, this remains a reasonable hypothesis that is not demonstrated yet. For this reason we preferred not to decorrelate this effect from the data and to leave it as an additional source of uncertainty.}

    \subsection{RF characterization}
    \label{sec_electromagnetic_measurements}

        In this section we present the measured electromagnetic performance of the QUBIC TD feed-horns compared with simulations obtained using the measured profiles of all the feed-horns in the array. In our measurements we tested a subset of feed-horns ($\sim$40\% for array 1 and $\sim$14\% for array 2), each identified with a pair of numbers corresponding to the row and column position in the array. 
    
\begin{figure}[h!]
        \begin{center}
            \includegraphics[width=15cm]{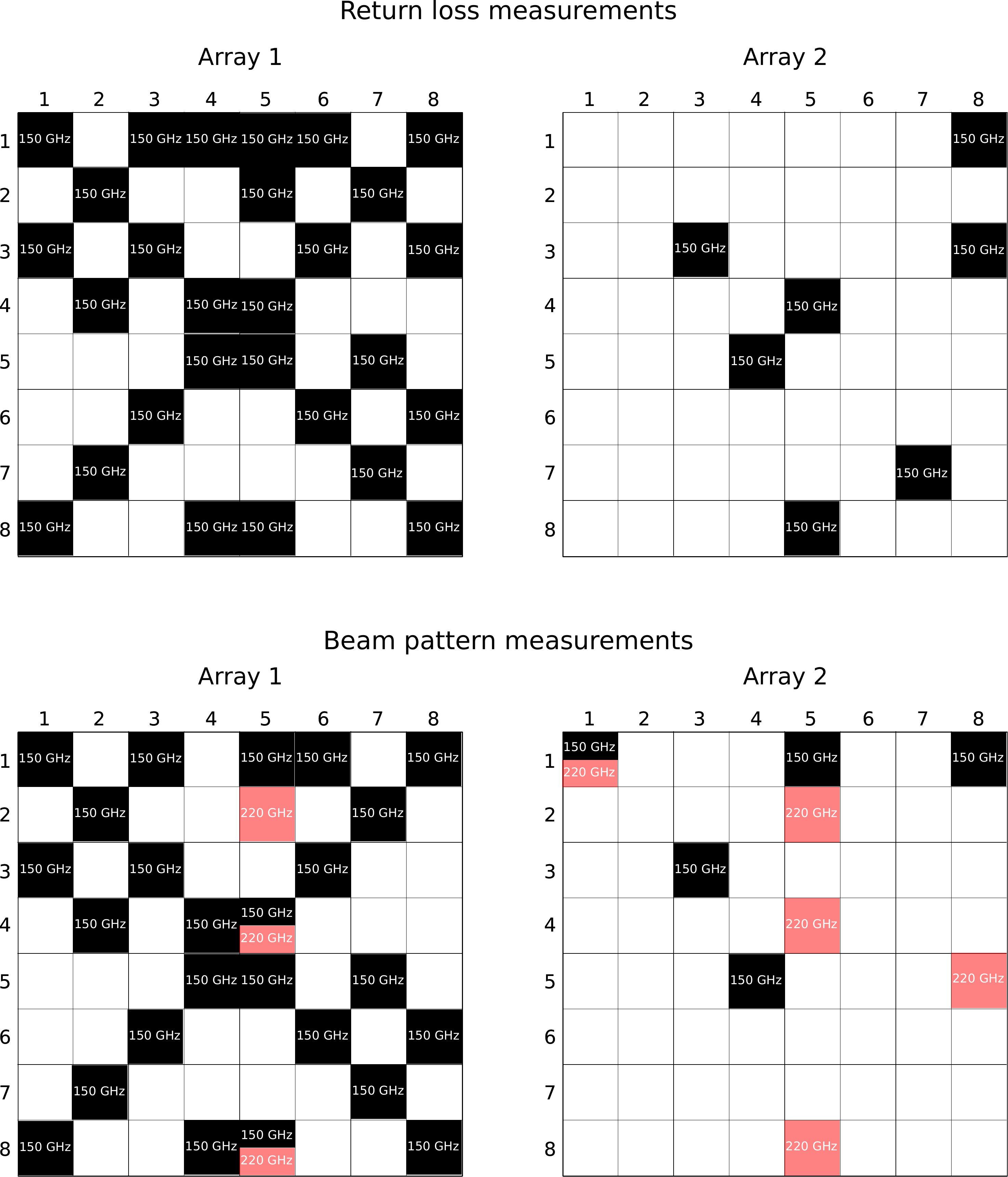}
        \end{center}
        \caption{\label{fig_measured_horns}Grids summarizing the electromagnetic measurements carried out on the TD feed-horn arrays. See text for further details.}
    \end{figure}

    In figure~\ref{fig_measured_horns} we show four grids summarizing the measurements that we have conducted on the TD arrays. Each grid represents the 8$\times$8 horn array in each module. On the top of each grid we have specified the number of the module and the type of measurements performed. In each grid we identify with black cells the horns tested at 150\,GHz, with red cells the horns tested at 220\,GHz and with black-red cells the horns tested at both frequencies. Notice that we do not report return loss measurements for the TD array at 220\,GHz. In this case the measurements presented in this paper were carried out on the two central horns of the two modules of the QUBIC FI that share the identical electromagnetic design and manufacturing technique of the TD (see section~\ref{sec_fi_horns}).
    
  \subsubsection{Experimental setup and procedures}
  \label{sec_setup_procedures}
    
\begin{figure}
        \begin{center}
            \includegraphics[width=7.0cm, height=5.3cm]{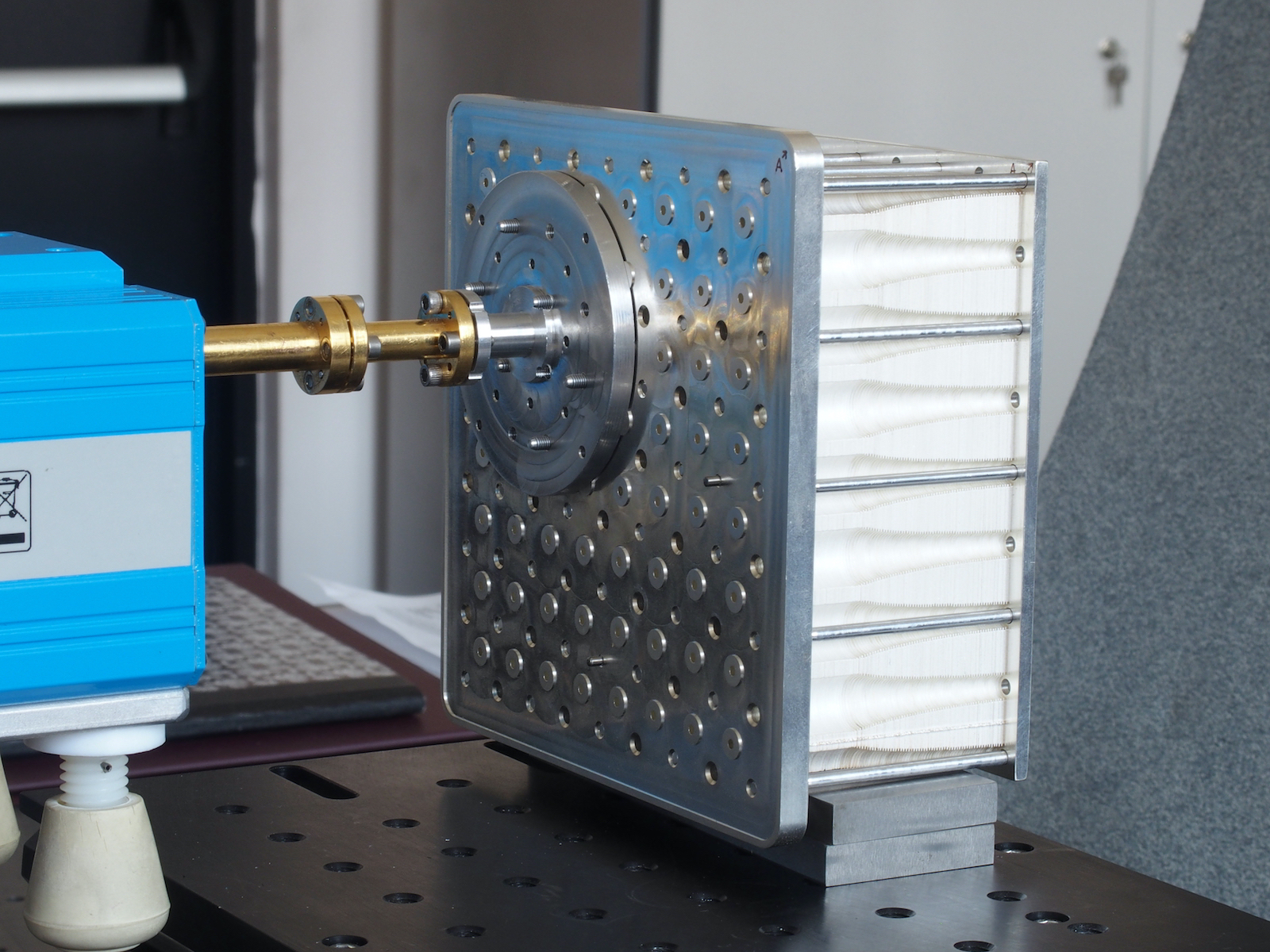}
            \includegraphics[width=7.0cm, height=5.3cm]{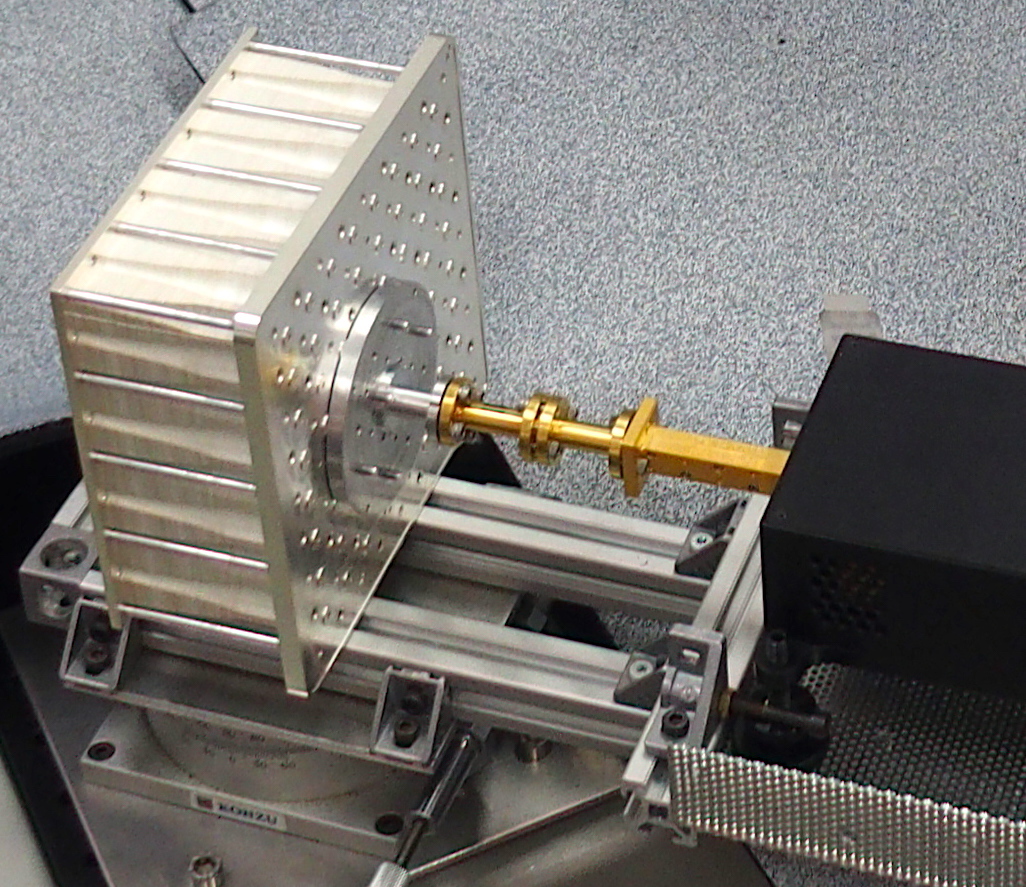}
        \end{center}
        \caption{\label{fig_em_measurements_setup1}\textit{Left}: detail of the adapter cascade used in the 110--170 GHz measurements. After the VNA golden millimeter extension where the TRL calibration plane is defined, there is a golden rectangular to circular \newb{waveguide} taper. The golden taper is connected to the horn array by means of silver flange adapter. \textit{Right}: detail of the adapter cascade used in the 170--260 GHz measurements. A rectangular taper from WR4.3 to WR06 is attached to the VNA golden \newb{waveguide} millimeter extension (TRL calibration plane). After that, the same components of the lower \newc{band} set-up is used.}
    \end{figure}

    \new{The experimental setup consisted of a vector network analyzer (VNA) equipped with millimeter extensions for full two-port characterization in the 110--170\,GHz and 170--260\,GHz bands. With this setup we measured the horn return loss and radiation patterns and compared our measurements with simulations. We connected the array to the VNA by means of a cascade of adapters (figure \ref{fig_em_measurements_setup1}) to connect the standard rectangular \newb{waveguide} of the VNA to the custom circular one of the horns.}
    
    \newc{The reference plane for the VNA measurements is at the WR06 millimeter-wave extension waveguide port and calibrated with thru-reflect-line (TRL) standards. The time domain window gating function of the network analyzer system is used to isolate the horn back-scattered signal from the undesired reflections associated with the test adapters.} Because the VNA setup is sensitive to only the \newe{dominant} propagation mode we could not test the multi-mode behavior of the system in the 220\,GHz band. Consequently also the results of the simulations displayed in figures~\ref{fig_horns_measured_return_loss}, \ref{fig_measured_patterns_150GHz} and \ref{fig_measured_patterns_220GHz} regard single-mode propagation.
    
    \newc{To sample the beam patterns, the TD feed-horn array was mounted on a goniometer fixed on an optical bench. The horn array was illuminated in the 110--170 GHz band by a corrugated circular standard gain horns manufactured by Radiometer Physics, while in the 170--250 GHz band it was illuminated by a smooth wall standard gain horn manufactured by Custom Microwave. In these measurements we moved the \newe{device under test (DUT)} in azimuth with an angular step of 1$^\circ$ and selected the proper reference plane ($E$-plane or $H$-plane) by properly rotating the launcher and the DUT.} 


  \subsubsection{Results}
  \label{sec_results}
  
    In this section we summarize the measured return loss and beam patters with simulations and show that we obtain an overall match within the uncertainties given by the mechanical differences among the horns.
    
\begin{figure}[h!]
        \begin{center}
            \includegraphics[width=14cm]{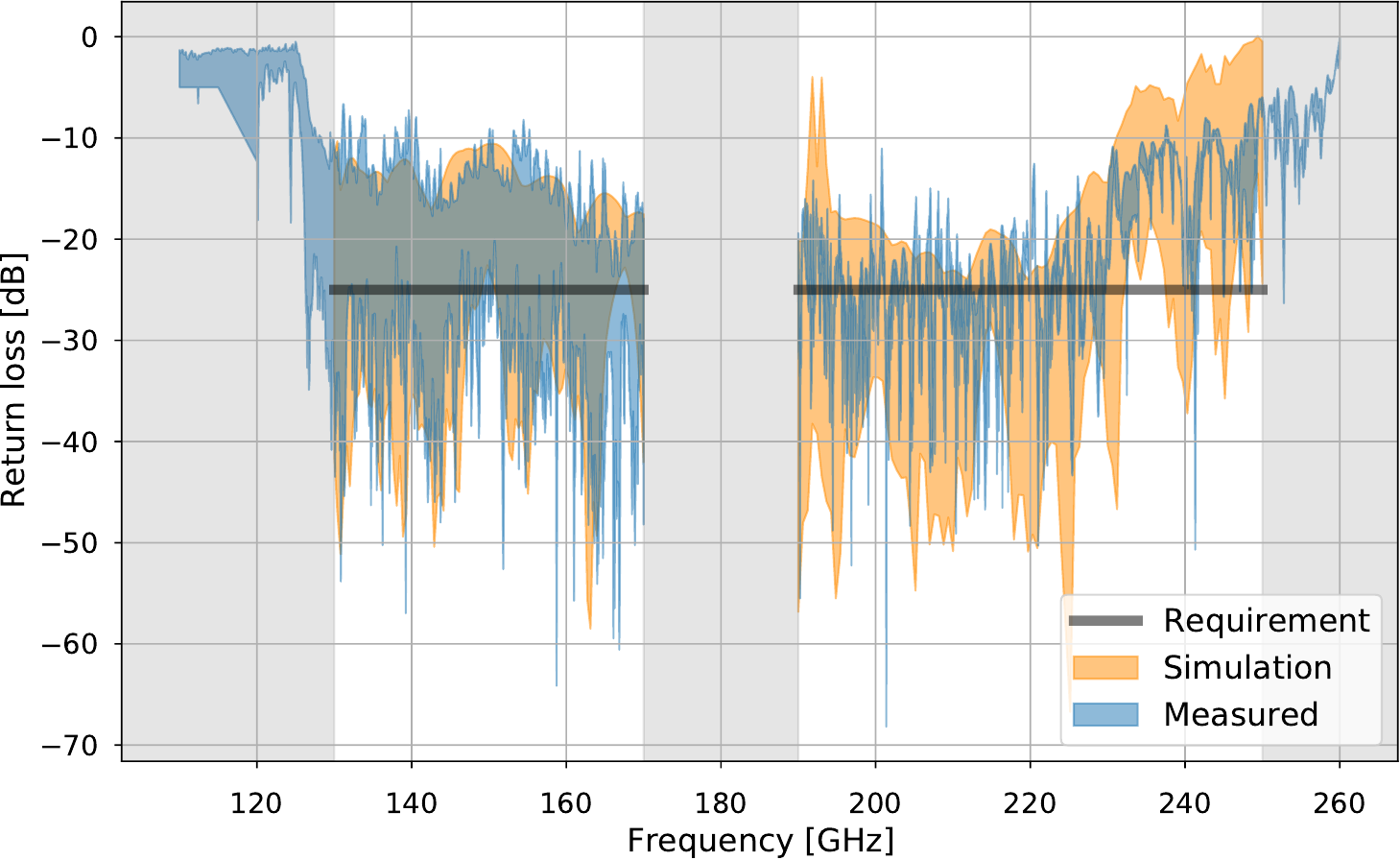}
        \end{center}
        \caption{\label{fig_horns_measured_return_loss}Measured return loss compared with simulation \newb{and requirement listed in table~\ref{tab_horn_requirements}}.}
    \end{figure}

	\paragraph{Return loss.} In figure~\ref{fig_horns_measured_return_loss} we show the results of the return loss measurement in both bands compared with the simulations. The orange area is the envelope of the return loss simulated for all the 128 feed-horns in the array, each with its own measured profile, while the blue area is the envelope of the measured return loss for all the tested horns (refer to figure~\ref{fig_horns_measured_return_loss}).
	
	We see that the measured reflection matches the simulation, within the scatter given by the mechanical differences among the horns. We also see that the average achieved return loss at 150\,GHz lies around $-$20\,dB, while in the higher band it is around $-$25\,dB up to 230\,GHz and then degrades to about $-$10\,dB as expected. \newb{We believe that the higher-than-expected return loss at 150\,GHz and} the scatter among simulations is caused by the out-of-spec in the mechanical tolerance discussed in section~\ref{sec_second_prototype_metrology}. Given the improvements adopted in the manufacturing procedure we \newb{think} that this scatter \newb{and the overall return loss are} reduced in the FI horn array. 
	
    The reader may notice that the measured return loss between 190 and 230\,GHz is about $-$25\,dB, therefore 5\,dB lower than the value resulting from the simulation of the nominal feed-horn (see the left panel of figure~\ref{fig_horn_simulated_return_loss_xpol}). This is because the measurements and the simulations displayed in figure~\ref{fig_horns_measured_return_loss} are relative to single-mode propagation, while the simulation in figure~\ref{fig_horn_simulated_return_loss_xpol} considers all the possible modes that can propagate in the 220\,GHz band.
    
\begin{figure}[h!]
        \begin{center}
            \includegraphics[width=14cm]{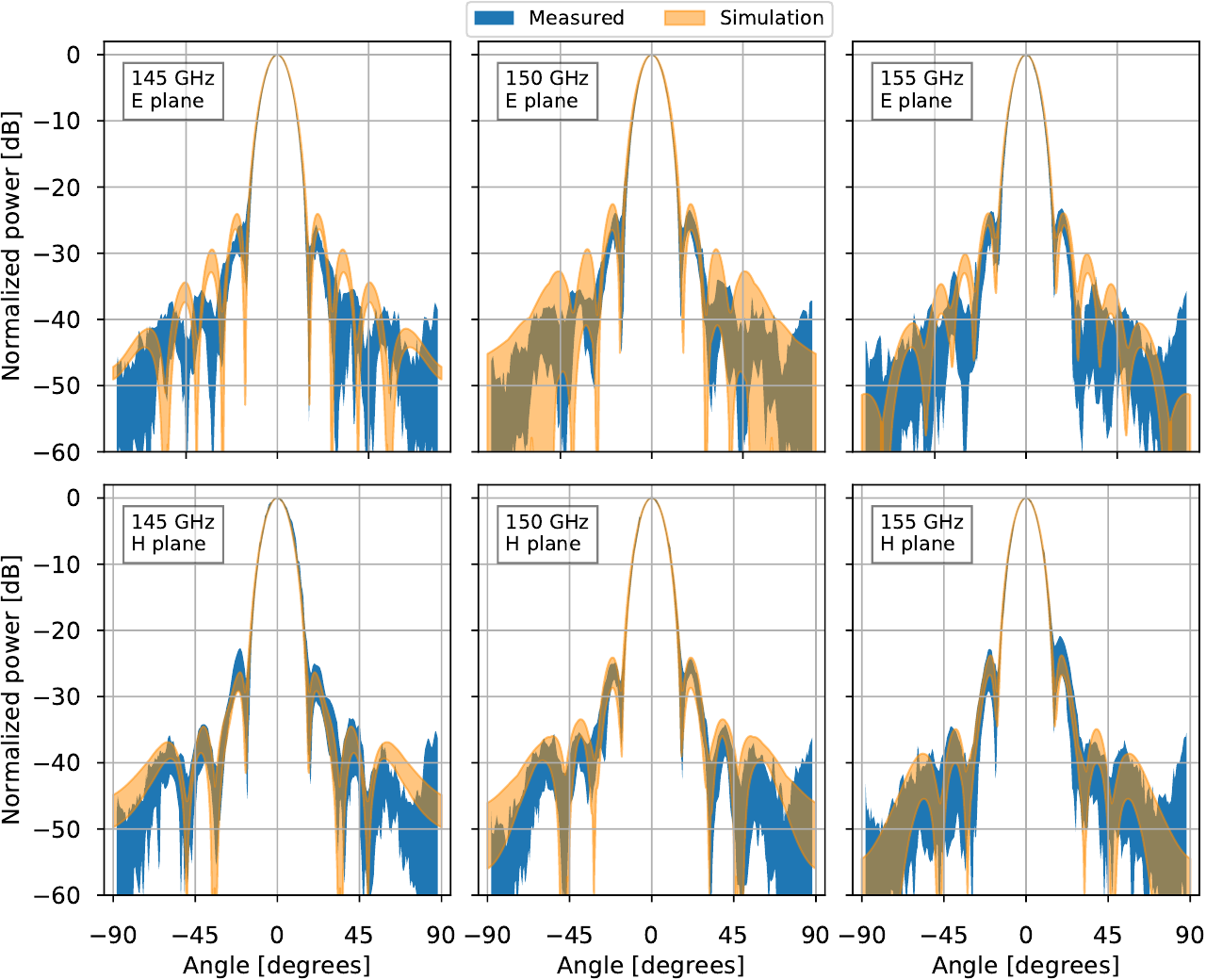}
            \caption{\label{fig_measured_patterns_150GHz}Measured co-polar beam patterns at 150\,GHz compared with simulations. \textit{Top:} $E$-plane. \textit{Bottom:} $H$-plane.}
        \end{center}
    \end{figure}

\begin{figure}[h!]
        \begin{center}
            \includegraphics[width=16cm]{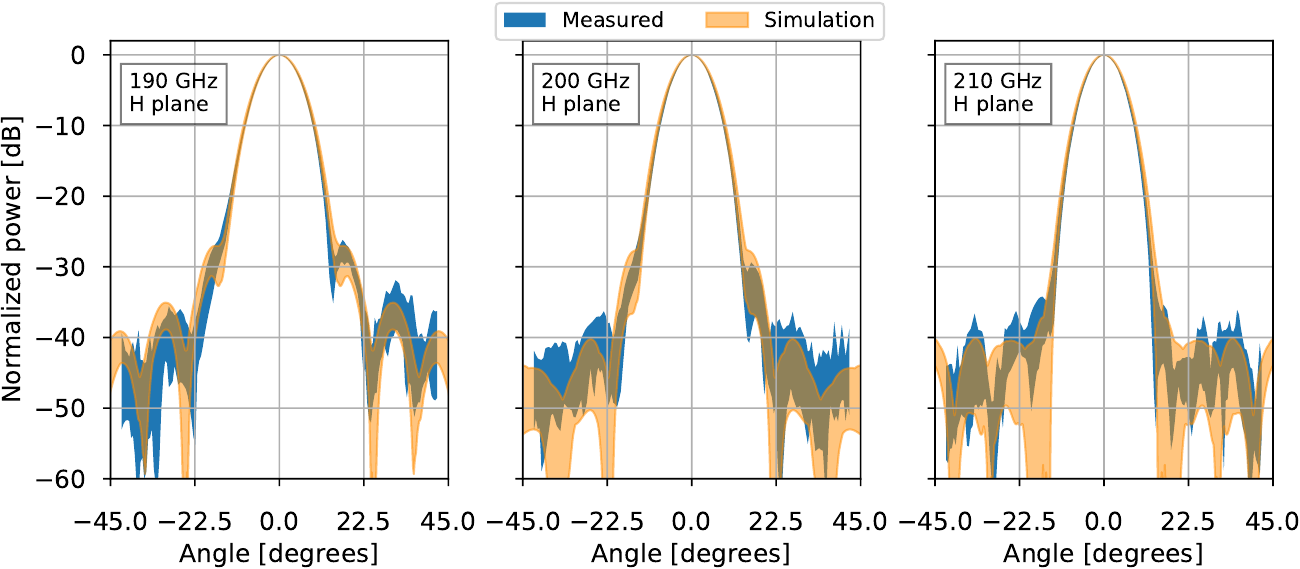}
            \mbox{}\\
            \includegraphics[width=10.67cm]{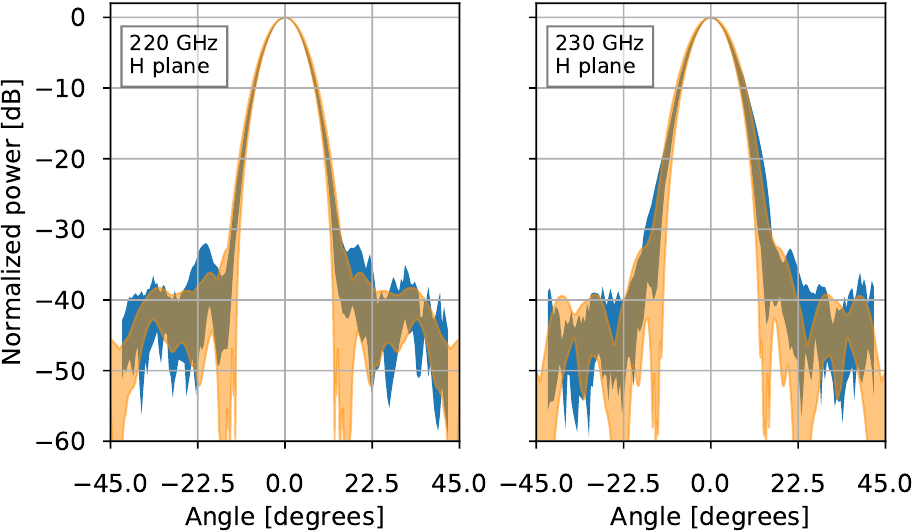}
            \caption{\label{fig_measured_patterns_220GHz}Measured co-polar ($H$-plane) beam patterns in the 220\,GHz frequency band.}
        \end{center}
    \end{figure}

    
    \paragraph{Beam patterns.} We show our beam pattern measurements compared with simulations in figures~\ref{fig_measured_patterns_150GHz} and \ref{fig_measured_patterns_220GHz}. Also in these figures the orange area is the envelope of the simulated patterns for all the feed-horns in the arrays and the blue area is the envelope of the measured patterns.
    
    In the 150\,GHz band (figure~\ref{fig_measured_patterns_150GHz}) we measured $E$- and $H$-plane diagrams in the entire D band (110-170 GHz) but for sake of simplicity reported them for only three frequencies: 145, 150 and 155\,GHz. Measurements match simulations very well (with a few dB discrepancy) down to about $-$30\,dB. The scatter increases at larger angles, where the detected power is smaller and the measurement becomes sensitive to signal reflections.
    
    We have obtained similar results in the 220\,GHz band (figure~\ref{fig_measured_patterns_220GHz}). In this case we reported only the $H$-plane diagram at five frequencies, equally spaced between 190 and 230\,GHz. Also in this band there is a very good match between measurements and simulations down to $-$30\,dB.
    
\begin{figure}[h!]
        \begin{center}
            \includegraphics[width=14cm]{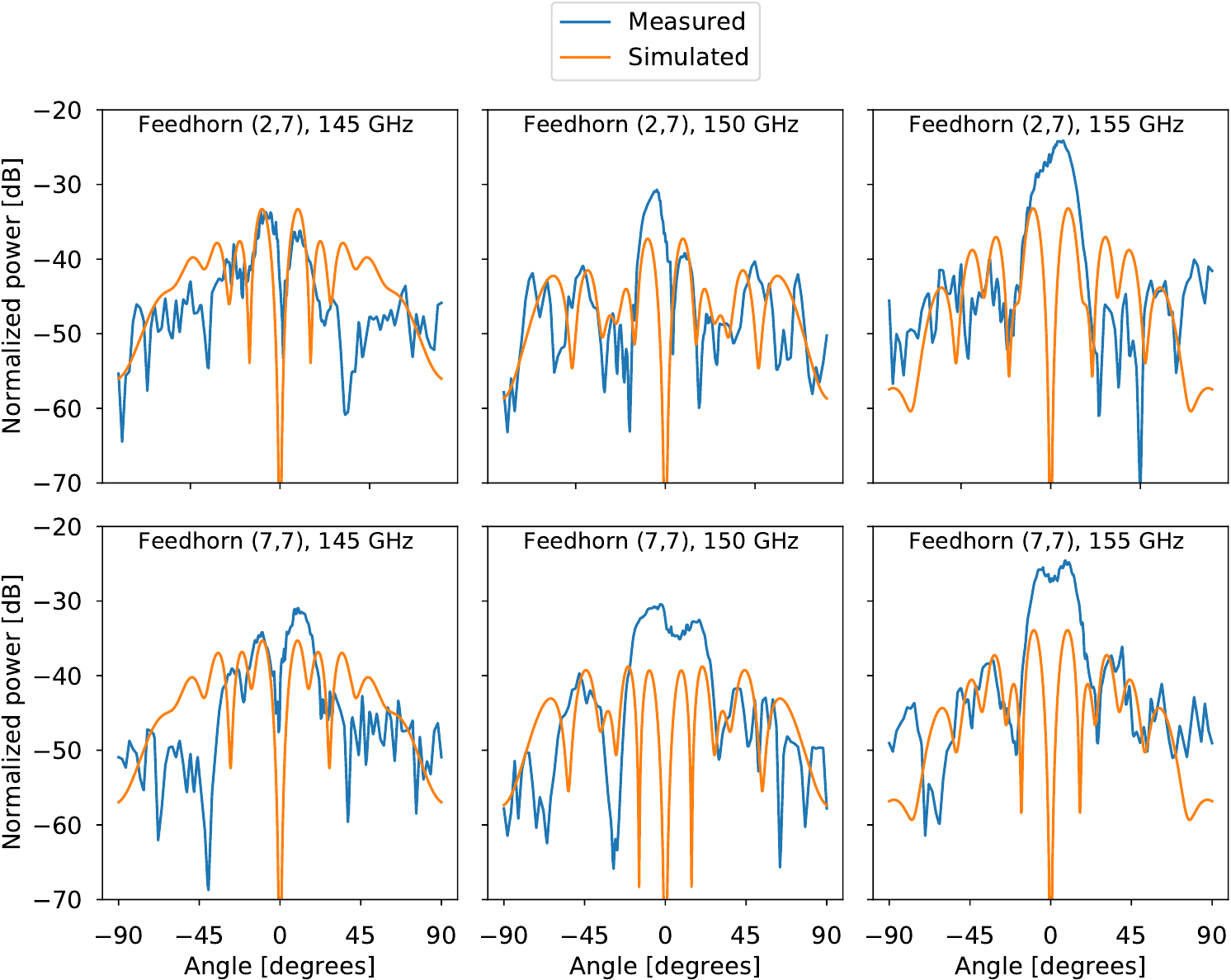}
            \caption{\label{fig_measured_patterns_xp150GHz}Measured and simulated cross-polar (plane $\phi= 45^\circ$) beam patterns in the 150\,GHz frequency band. \newb{The patterns are normalized to the 0\,dB maximum of the co-polar patterns.}}
        \end{center}
    \end{figure}

    \new{We have also measured the cross-polar beam pattern in the 150\,GHz band for two horns of the array 1. We show these patterns in figure~\ref{fig_measured_patterns_xp150GHz}, where the measured data are compared with simulations. We see that measurements and simulations match at the level of the pattern general level and trend, although our setup was not optimized against reflections to allow a fine-grained match. This is particularly apparent at the highest frequency, where we see a mismatch of about 10\,dB in the maximum cross-polarization.}
    
    \subsection{The full instrument feed horn array}
    \label{sec_fi_horns}
    
\begin{figure}[h!]
    \begin{center}
        \includegraphics[width=14cm]{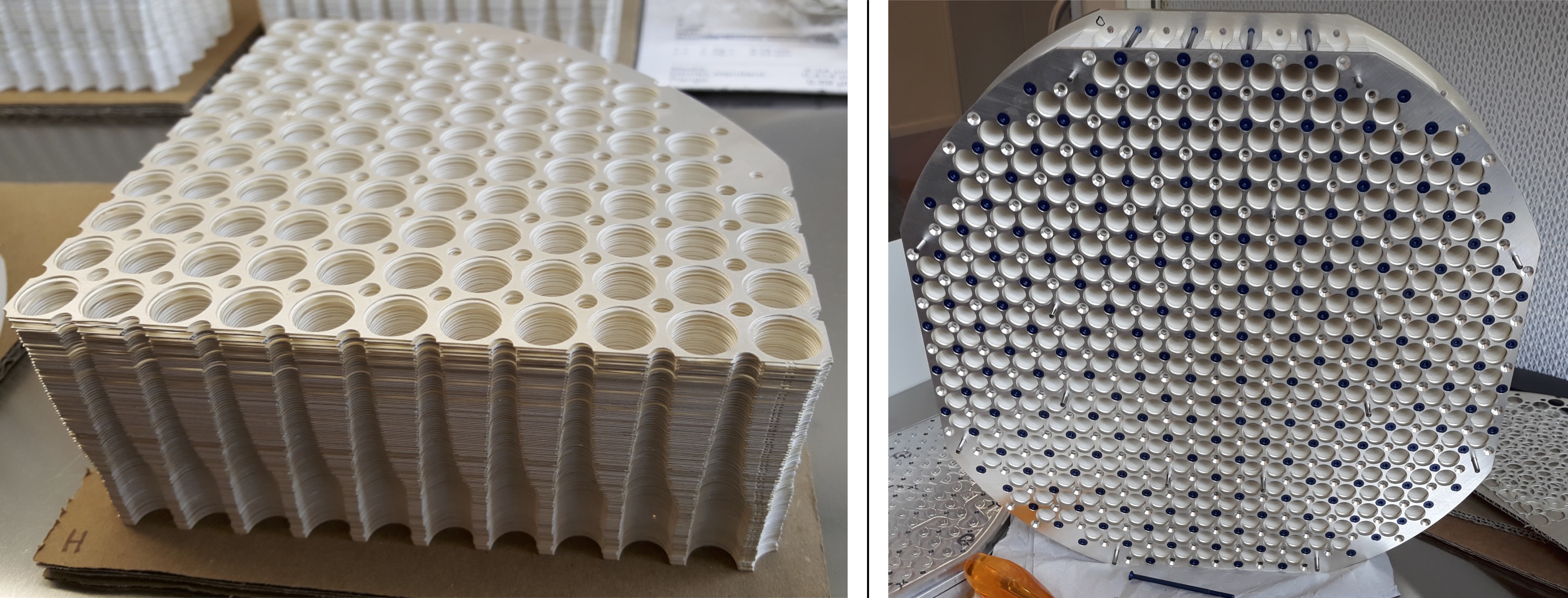}
        \caption{\label{fig_fi_horn_array}The back-to-back feed-horn system of the QUBIC FI. \textit{Left:} Stacked platelets of a quarter array. \textit{Right:} one of the two 400 horn arrays.}
    \end{center}
\end{figure}

        Here we present briefly the feed-horn array developed for the QUBIC \new{FI}, shown in figure~\ref{fig_fi_horn_array}. The FI horn system has the same electromagnetic design of the TD and it was manufactured with the same technique: the inner part by chemically etching 0.3\,mm thick aluminum sheets and the front and back flanges by mechanical milling 3\,mm and 6\,mm thick aluminum plates. \new{Like for the TD array we silver-coated all plates before the final integration}.

In this array we measured the geometrical profile of a large subset of horns in the two modules. This allowed us to simulate the expected performance of this subset and to compare it with laboratory measurements.

The detailed discussion of the development and testing of this part is out of the scope of this paper so that we defer the full discussion of the QUBIC FI horns to a forthcoming dedicated paper.

\section{Switch system}
\label{sec_switches}


    \subsection{Switch requirements and design}
    \label{sec_switches_requirements_design}

        The QUBIC switch array is used to select the baselines under test during the calibration phase. Theoretically the self-calibration procedure requires to acquire data from a known calibration source for all possible baselines, each obtained by closing all the horns apart from the pair of the selected baseline. \new{This approach, however, would cause a huge variation of the radiative load unto the focal plane, shifting the detectors' temperature well below the design one.} In QUBIC we apply a different, but mathematically equivalent approach \newc{\cite{Bigot-Sazy2013}} in which we close only the horns of the baseline under test, thus maximizing the signal stability during calibration.

In table~\ref{tab_witch_requirements} we list the main requirements of our switch system. In particular the insertion loss must be as low as possible, even if the 1\,K working temperature helps at least in reducing its thermal noise, while we require high isolation between on and off states and fast commutation for accurate and efficient baseline calibration.

\begin{table}[h!]
    \renewcommand{\arraystretch}{1.5}
    \caption{\label{tab_witch_requirements}Main requirements for the QUBIC switch array at room temperature}
    \begin{center}
        \begin{tabular}{m{2cm} c m{4cm}}
            \hline
            Requirement& Value & Notes \\
            \hline
            \hline
            Insertion loss\dotfill & $< -0.1$\,dB & Driven by minimization of signal loss and thermal noise. This is not met by the TD which is a factor 2 worse. \\
            Isolation\dotfill & $> 50$\,dB & Driven by maximizing contrast in fringe patterns during calibration.\\
            Return loss\dotfill & $<-20$\,dB & Over the 130--240\,GHz bandwidth.\\
            Switch commutation time\dotfill & $\sim$\, ms & To be negligible in the calibration duty cycle.\\
            Heat load\dotfill & as low as possible & To minimize thermal drifts of the 1\,K \newb{stage}.\\
            Mass\dotfill & as low as possible & Must be suspended on the top of the optical combiner.\\
            \hline
         \end{tabular}
    \end{center}
\end{table}

\begin{figure}[h!]
        \begin{center}
            \includegraphics[width=12cm]{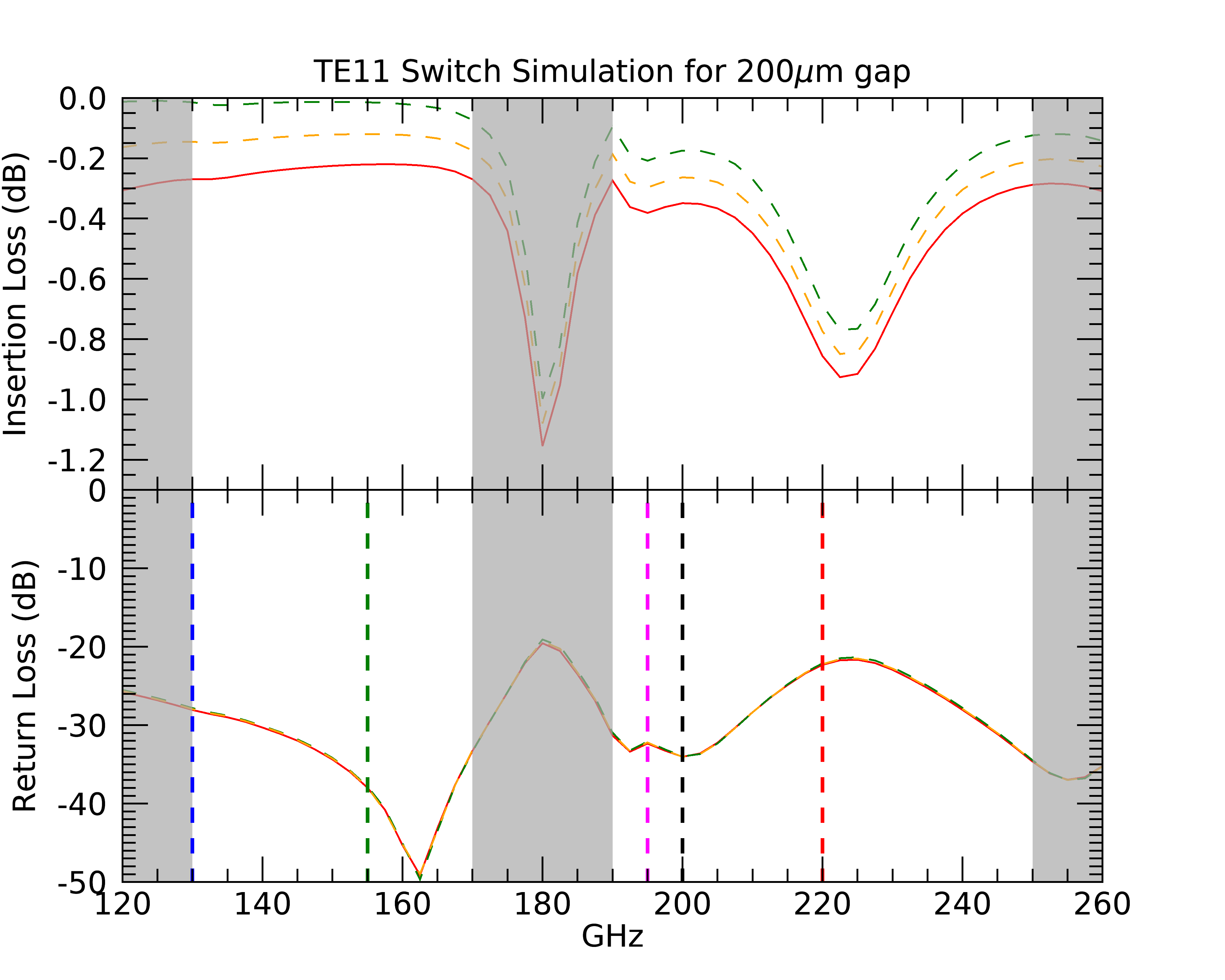}
        \end{center}
        \caption{\label{fig_TE11_switch_simulation}\newd{\textit{Top}: Simulation of TE11 mode insertion loss of the switch waveguide for a choke gap of 200 $\mu$m. The red solid line is for an effective resistivity $\rho=16~\mu \Omega\,\mathrm{cm}$, the green dashed line is for a perfect conductor ($\rho=10^{-6}~\mu\Omega\,\mathrm{cm}$ for computational reasons) and the orange dashed line is for Al-6061 resistivity at room temperature ($\rho=4.19~\mu\Omega\,\mathrm{cm}$). \textit{Bottom}: Simulation of TE11 mode return loss of the switch waveguide for a choke gap of 200 $\mu$m. Since the return loss is dominated by the switch geometry, the cases for the three resistivities ($\rho=[0,4.19,16]~\mu\Omega\,\mathrm{cm}$) are practically identical. The dashed vertical lines are placed where the higher order modes start to propagate. We used the same color code as in figure \ref{fig_S11_Choke_High_order_modes}.}}
    \end{figure}

From an electromagnetic point of view, the switch design was based on a best effort approach where the length of the \newb{waveguide} is set by the shutter mechanism dimensions plus the thickness of the plates housing the threaded holes to mate the two horns arrays. The \newb{waveguide} diameter matches the input \newb{waveguide} of the horn. The switch \newb{waveguide} was designed with a choke trap to short-circuit the gap when the switch is open. Such a structure was originally designed for the lower band, but extended simulations showed that it was effective also for the upper band. In figure~\ref{fig_TE11_switch_simulation} we report the results of the simulation of the transmission and reflection of the main electromagnetic mode (TE11) for 25\,mm \newb{waveguide} length and $200\,\mu$m choke gap. In the two QUBIC bands the reflection is always below $-$20\,dB. 

\newd{We have simulated the insertion loss at room temperature with an effective bulk resistivity of 16\,$\mu\Omega\,\mathrm{cm}$, which takes into account the effect of surface roughness (O. Peverini, private communication). The simulated loss (reported in figure~\ref{fig_TE11_switch_simulation}) is a factor two larger than the requirement, although it is consistent with the loss of a waveguide of identical length without the choke ($-$0.18\,dB). For comparison, an idealized loss-less waveguide (actually $\rho=10^{-6}\mu\Omega \,\mathrm{cm}$ for computational reasons) with the same gap and choke has an insertion loss of $-$0.02\,dB. \newe{This is the reactive mis-match loss associated with the junction geometry.} We have also simulated the insertion loss for an ideally smooth waveguide using the bulk resistivity of Al6061-T6 ($\rho=4.19~\mu\Omega\,\mathrm{cm}$) as reported in Clark et al. \cite{CLARK1970295} and in ASM HANDBOOK \cite{Handbook...Al...1990}. Since the return loss is dominated by the switch geometry, the output of the simulation for the three resistivities are, in this case, practically identical.} 

At higher frequencies the loss is definitely higher than desired. This can be partially due to the almost double length of the \newb{waveguide}, if measured in wavelength units. Another reason why the loss is higher in the upper band is because the choke itself is less effective. \newc{While in the lower band the discontinuity of the waveguide is made invisible to the incoming radiation by a choke trap which lowers the reflected power at the gap below $-40$ dB (for the waveguide and choke geometry see figure \ref{switch_signle_channel_prototype}), in the higher band the reflected power rises at $-12$dB.} Moreover, starting from 190\,GHz TE11 begins to convert to higher order modes and the simulation is less revealing.

\begin{figure}[h!]
        \begin{center}
            \includegraphics[width=12cm]{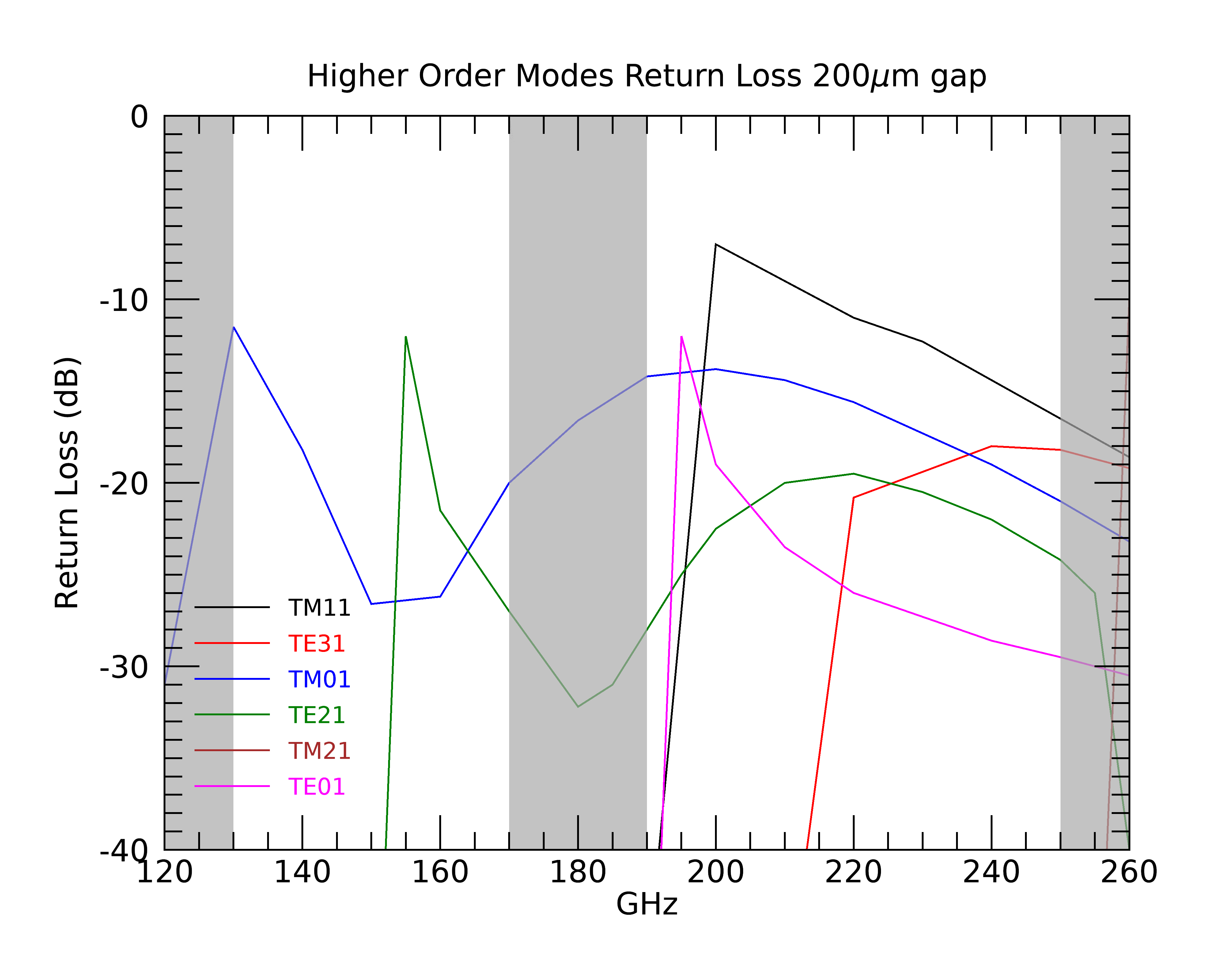}
        \end{center}
        \caption{\label{fig_S11_Choke_High_order_modes} Simulation of the return loss induced by the $200\mu$m gap where the shutter enter to short circuit the \newb{waveguide}.}
    \end{figure}

\new{In order to verify that the higher modes are not reflected back by the \newb{waveguide} gap, we run a simulation of the return loss including all the  modes able to propagate within our band, i.e. from the TM01 up to the TM21 mode. The result, reported in figure \ref{fig_S11_Choke_High_order_modes}, shows that on average this return loss is less than 10\%.} 

\begin{figure}
        \begin{center}
            \includegraphics[width=7.21cm]{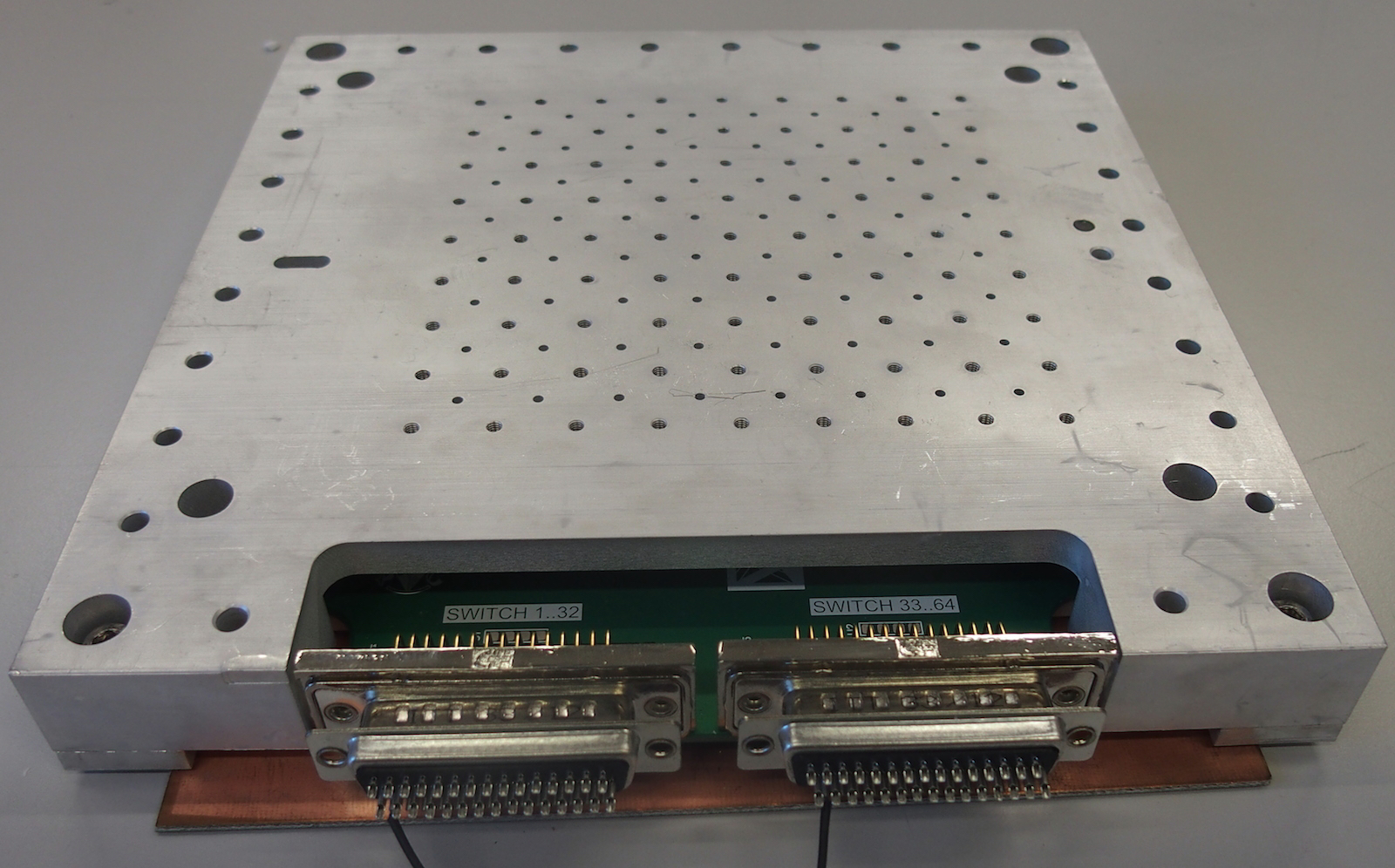}
            \includegraphics[width=7.13cm]{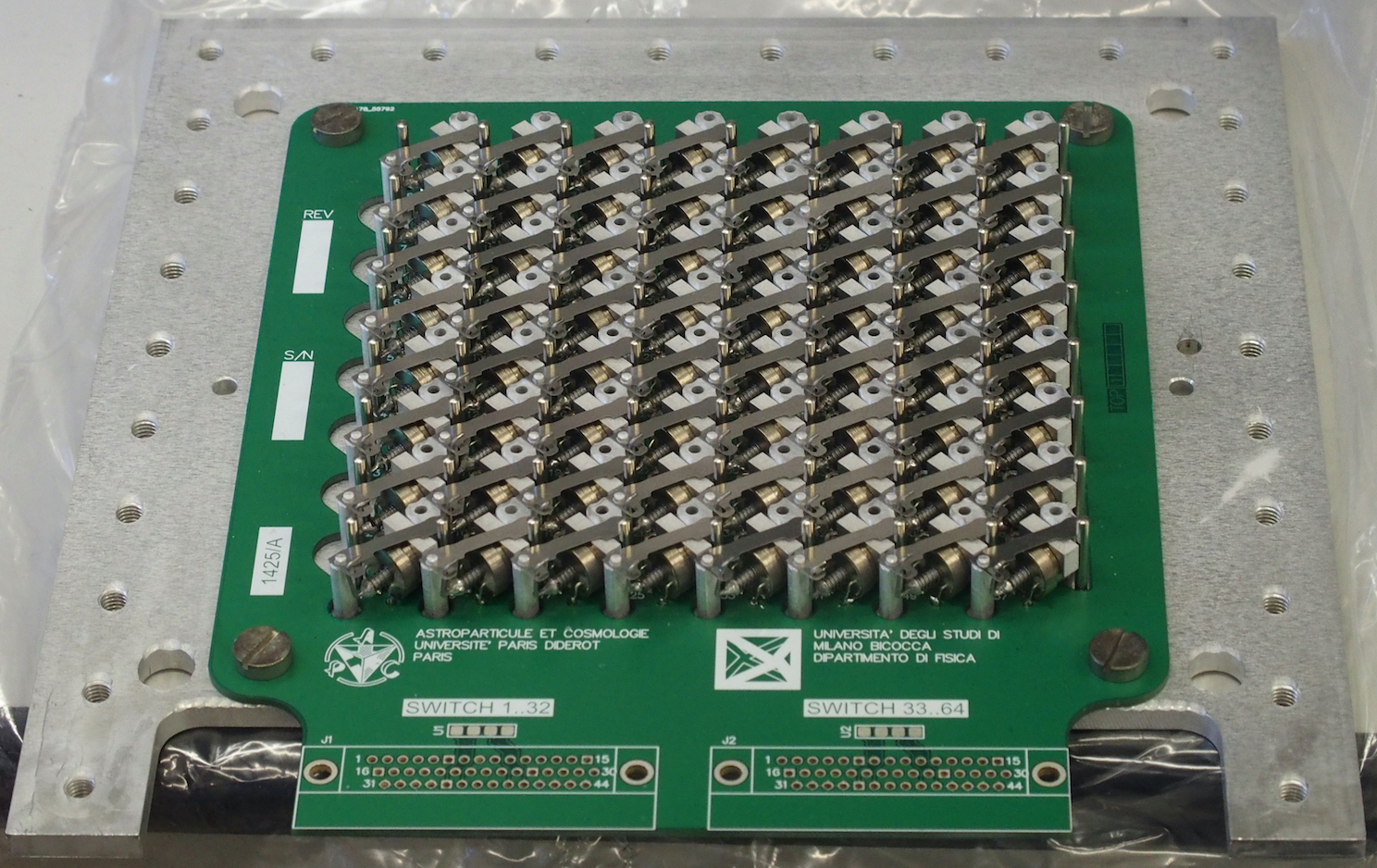}
        \end{center}
        \caption{\label{fig_TD_switch}\textit{Left}: the switch array closed and ready to be integrated with the horn arrays. \textit{Right}: the inner part of the switch assembly with solenoids, shutters and waveguides clearly visible.}
    \end{figure}

\begin{figure}[h!]
        \begin{center}
            \includegraphics[width=7.0cm]{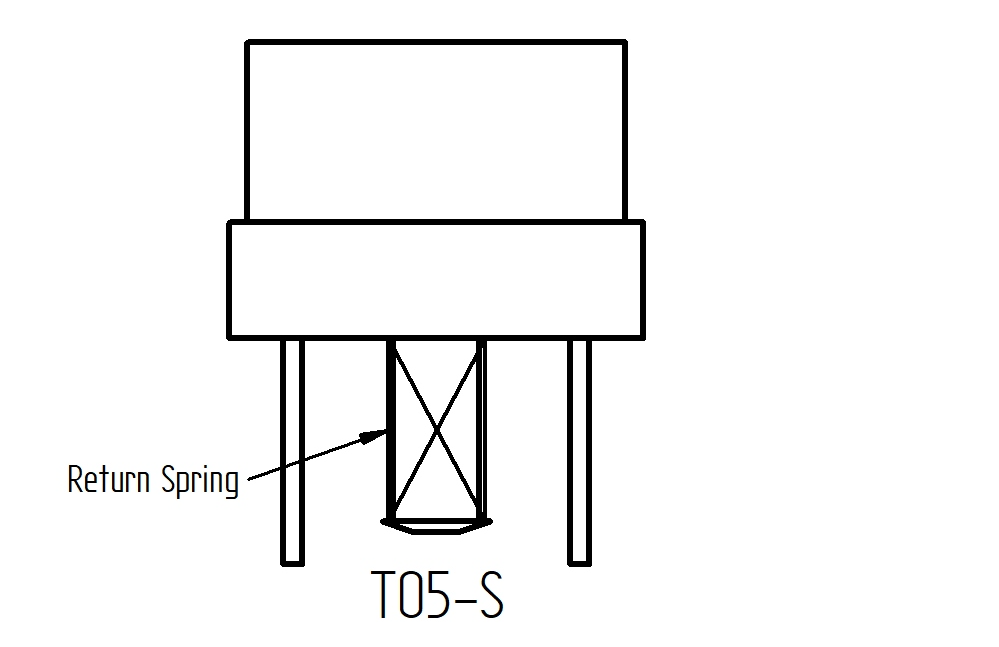}
            \includegraphics[width=6.0cm]{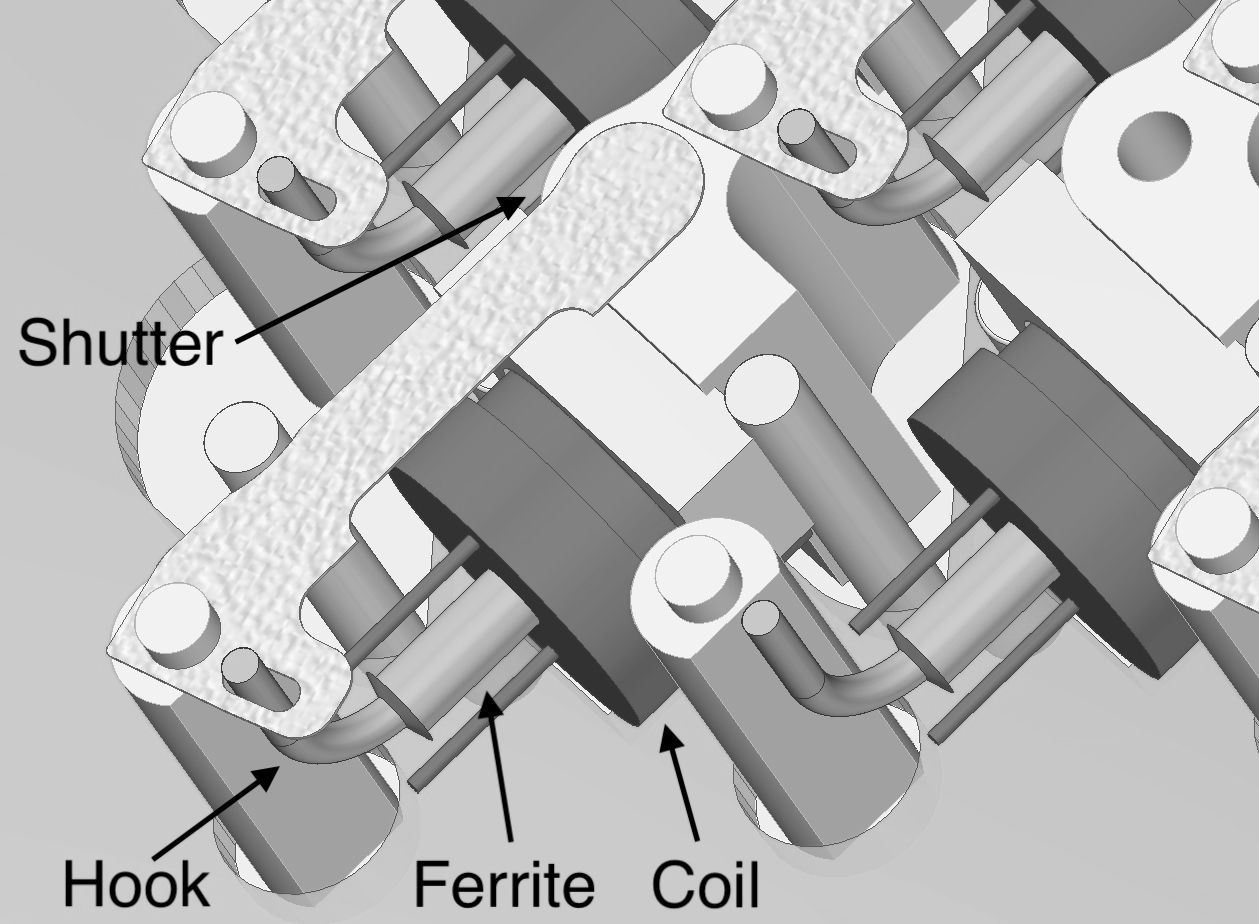}
        \end{center}
        \caption{\label{fig_single_switch_elements}\textit{Left}: the TO5-S coil used to move the shutter inside and outside the \newb{waveguide}. \textit{Right}: detail of the 3D model where single components of a unit cell are identified.}
    \end{figure}

\new{Since QUBIC is designed to observe the low/medium $\ell$ multipoles, the horns' spacing is one of the smallest possible. As a matter of fact, a triangular lattice is even more compact, but it would have left no room for the platelets tightening screws and for the shutters' mechanism. The QUBIC's square lattice is a compromise that leaves enough room available to realize the mechanical shutters (switches). We designed each switch around the smallest commercially available coil manufactured by Line Electric (model TO-5S figure \ref{fig_single_switch_elements} \textit{left}). The area occupied by the whole device is so small that the minimum center-to-center distance between horns is not determined by the switch, but by the horn mouth diameter. The total thickness of the switch block is 24.75 mm and this is also the \newb{waveguide} length. The TD switch array in its final form can be seen in figure \ref{fig_TD_switch}, with and without the lid assembled. Top panel shows the array of newb{waveguides} and threaded holes to fix the top horn array. The two DB-37 connectors are used to bias the coils. Bottom panel shows the internal part of the switch array. All the coils, shutters and \newb{waveguides} are clearly visible.}

\new{The shutter is realized by means of a thin stainless steel blade entering the \newb{waveguide} through a thin gap and acting as a short circuit. The gap is 200 $\mu m$ wide and the blade is 100 $\mu m$. The back and forth blade movement is actuated by the coil pulling a ferrite which is soldered to a hook attached to the shutter. The shutter makes an angular movement around a pivot. The shutter returns to its original position thanks to a spring pushing back the ferrite. The center of figure \ref{fig_single_switch_elements} shows a detail of the TD solid model, with its main components highlighted. On the right of the same figure, a microscope picture of the device is reported.}

\new{The coil nominal ohmic impedance is 80\,\ohm\ at room temperature, but lowers to 19\,\ohm\ at 4\,K.} We use a constant current of 90\,mA to hold the ferrite inside the coil, resulting in a Joule dissipation of 150\,mW per shutter. As already stated, we decided to limit to two the number of switches closed at the same time. However, from a thermal point of view the system can manage four switches closed at the same time. As a matter of fact, if we limit to four the number of simultaneously biased switches we reduce the extra heat load to 600\,mW. Thanks to the large mass of the horns plus switches system, the heat released increases horn-switch block temperature on a timescale much longer than the RF load change without any impact on the self-calibration effectiveness. As a matter of fact, during the test of the assembled TD, we found that the heat released increased horn-switch block temperature by less than 2\,mK in a timescale of 2 minutes. This effect is not detected by the focal-plane detectors because of \new{a Neutral Density (ND) filter located on the 1-K cold stop, as described in fig.\ref{fig_intrument_system_overview} caption. This ND filter was added for the testing campaign in the lab in order to reduce the background power on the detectors. When observing the sky the ND filter will be removed.}

Another reason not to exceed the number of four active switches (two per electronic module, which are two in the TD) is because the minimum number of wires necessary to drive $N$ devices is $(N+1)$, where the extra wire is used for current return of the $N$ coils. Since the driving current per coil is not negligible, the return wire must be made of copper instead of phosphor-bronze or manganine (this is discussed in detail in section~\ref{sec_TD_switch_manufacturing_and_control_electronics}). To minimize the conductive heat load, the copper wire must be thin ($\leq 200\,\mu$m), reducing the number of coils active at the same time to two per return wire. Since there are two switch banks, operated by two electronic boards, there are a maximum of four switch shut at the same time. However, since the self calibration procedure is effective even with only two horns closed, we limit to two the active switches operated at the same time.

	\subsection{Single channel prototype}
	\label{sec_single_channel_prototype}

\begin{figure}[h!]
    \begin{center}
        \includegraphics[width=14cm]{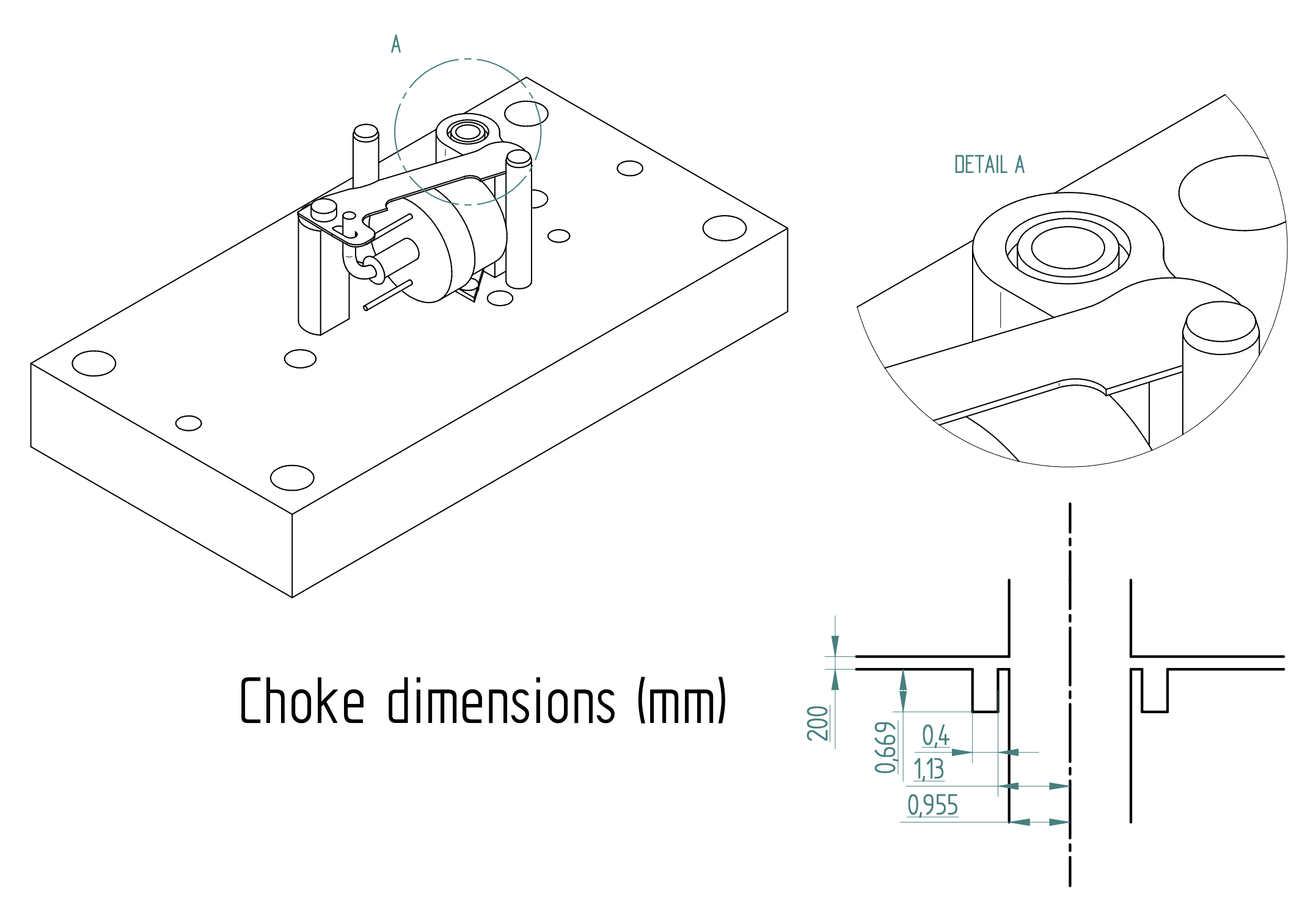}
        \caption{\label{switch_signle_channel_prototype} \new{Solid model of the single channel prototype: the lid was removed to reveal the solenoid and the shutter in their open position. The detail magnifies the \newb{waveguide} with the choke trap. A schematic cross section of the \newb{waveguide} and choke with the drawing dimension is shown too.}}
    \end{center}
\end{figure}

We developed a single channel prototype at the APC Laboratory in Paris \newd{\cite{Bordier2014}} to test the electromagnetic performance, the effectiveness of the linear motion design and the compatibility with the cryogenic environment. We realized both the single channel prototype and the full array for the QUBIC TD in Al-6061T6. \new{The 3-D sketch in figure~\ref{switch_signle_channel_prototype} shows the single channel switch without the lid to highlight the coil pulling the shutter inside a 200\,$\mu$m gap, in between the two sections of the circular \newb{waveguide}. Details of the choke trap are also given. When the shutter is outside the \newb{waveguide}, the switch is on and the coil is not biased. The shutter movements are limited by a couple of stainless steel alignment pins.}

\new{The prototype was tested four times in liquid nitrogen (LN2) at 77\,K to verify the coil lifetime. In each thermal cycle the switch worked at 50 Hz, accumulating hundred thousands operations. The functionality of the back and forth movement of the shutter was verified by means of an optical fiber. This prototype never showed any issue in LN2.}

	    \subsubsection{Switch electromagnetic measurements}
         \label{sec_switch_electromagnetic_measurements}
            
            We characterized the single channel prototype at room temperature in both the QUBIC bands: 135--165\,GHz and 190--220\,GHz using the VNA available at the laboratory of the University of Milano-Bicocca. We used a pair of rectangular-to-circular transitions to connect the VNA to the switch. Since the expected insertion loss of the switch was low, we measured and subsequently removed the contribution of the tapered transitions.

\begin{figure}[h!]
    \begin{center}
        \includegraphics[width=12cm]{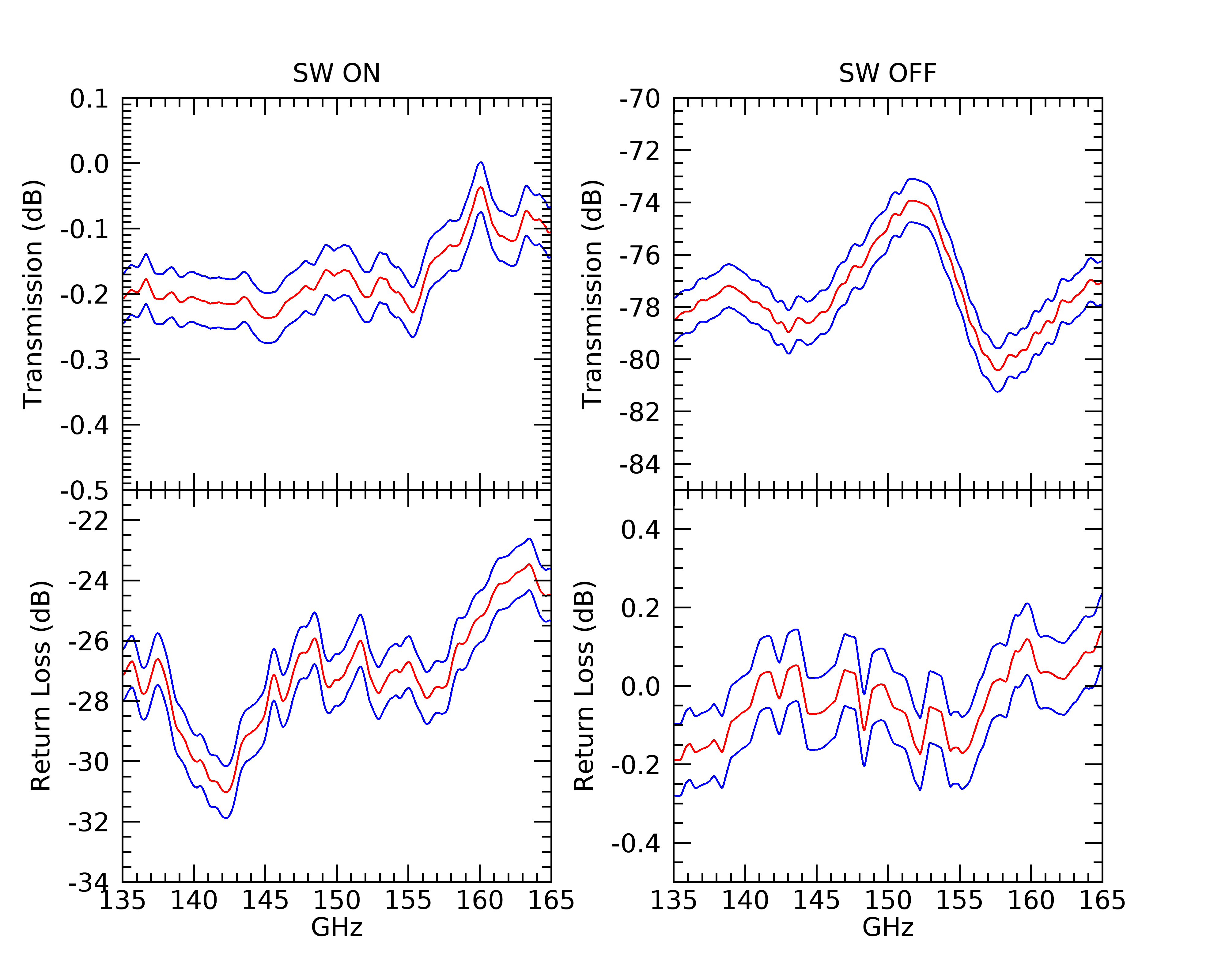}
        \caption{\label{fig_measured_scattering_param_protoype_135_165}Measured insertion loss (top) and return loss (bottom) of the switch prototype in the band 135--165\,GHz. \textit{Left:} Switch ON. \textit{Right:} Switch OFF. 1-$\sigma$ confidence bands are plotted in blue.}
    \end{center}
\end{figure}
\begin{figure}[h!]
    \begin{center}
        \includegraphics[width=12cm]{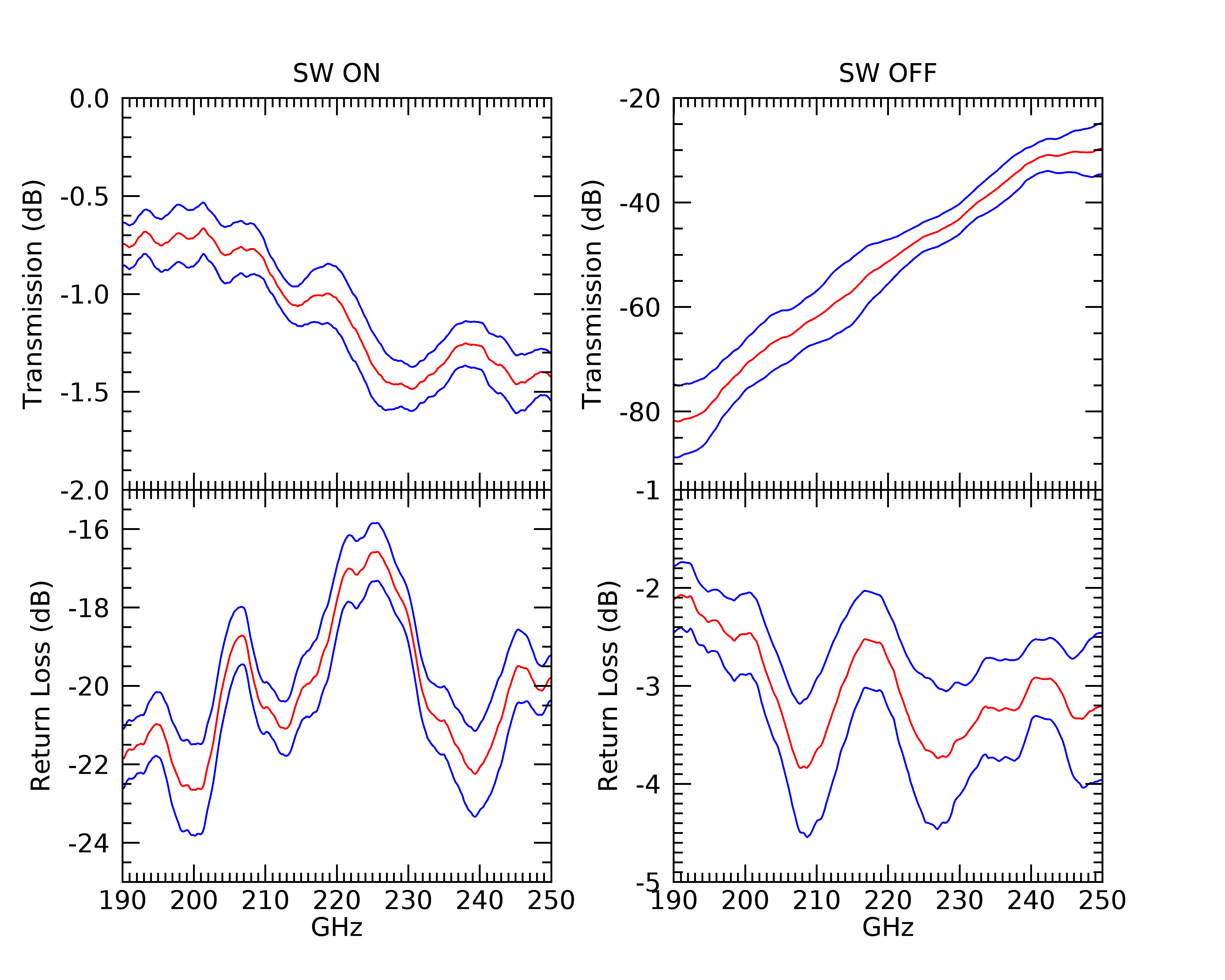}
        \caption{\label{fig_measured_scattering_param_protoype_190_220}Measured insertion loss (top) and return loss (bottom) of the switch prototype in the band 190--250\,GHz. \textit{Left:} Switch ON. \textit{Right:} Switch OFF. 1-$\sigma$ confidence bands are plotted in blue.}
    \end{center}
\end{figure}

\newd{When the switch is open, the insertion loss at 150\,GHz is $\mathrm{IL}=-0.2\pm0.05$\,dB at room temperature; see figures~\ref{fig_measured_scattering_param_protoype_135_165} and \ref{fig_measured_scattering_param_protoype_190_220} for which left panels report the insertion and return loss in the QUBIC lower band and the right panels report the same quantities for the upper band. The ohmic contribution to the loss is anticipated to be a monotonically increasing function of the guide surface roughness relative to the skin depth, which are $\sim 1\,\mu$m and 0.27\,$\mu$m (computed for $\rho=4.19~ \mu\Omega\,\mathrm{cm}$), respectively, at 150\,GHz \cite{1949JAP....20..352M,1992IMGWL...2..180C}. As already said in Section \ref{sec_switches_requirements_design}, the surface roughness associated with machining can be phenomenologically accounted for using an effective bulk resistivity of $16\,\mu\Omega\,\mathrm{cm}$ (O. Peverini, private communication) instead of the more conventional $\sim 4.19\,\mu\Omega\,\mathrm{cm}$ value for room temperature Al 6061-T6 \cite{CLARK1970295, Handbook...Al...1990}}. 

As reported in \ref{sec_switches_requirements_design}, there is a good agreement between simulations and measurements of the insertion loss at room temperature (IL=-0.2 dB at 150 GHz) while the simulation above 190 GHz can be inaccurate because of mode conversion from TE11 to TM11 (compare figure \ref{fig_TE11_switch_simulation} with figure \ref{fig_measured_scattering_param_protoype_190_220}). 
\newe{After cooling, the ohmic contribution to the insertion loss decreases and the contribution associated with surface roughness mode conversion remains practically unchanged.} \newd{Since the Al 6061-T6 resistivity decreases from 4.19\,$\mu\Omega\,\mathrm{cm}$ at room temperature down to 1.38 $\mu\Omega\,\mathrm{cm}$ at 10\,K \cite{CLARK1970295, Handbook...Al...1990}, we can estimate the insertion loss at cryogenic temperature to be $-$0.18\,dB at 150\,GHz and $-$0.73\,dB at 220\,GHz.} 
It is reasonable that also impedance mismatch and mode conversion are contributors to the insertion loss, but at the moment we don't have a way to estimate their contribution. 

The measured return loss, shown in the left bottom plot of figure \ref{fig_measured_scattering_param_protoype_135_165}, confirms that reflections are well within the requirement of $-$20\,dB as specified in table~\ref{tab_witch_requirements}.
At 220\,GHz the insertion and return loss requirements are not met, but this was expected because the prototype was designed and manufactured taking into account only the QUBIC lower band. Moreover, the TD works at 150\,GHz, so the 220\,GHz performance is not critical. \newc{In the FI the waveguides will be manufactured adopting all the techniques able to reduce the surface roughness like a combination of electrical discharge milling, high-grade reamers and chemical polishing to approach the desired electromagnetic specifications. In case a cryogenic characterization will be crucial, we will adopt the setup described in \newd{Zannoni et al.} \newc{\cite{2012SPIE.8452E..2RZ}}.}

    \subsection{The technological demonstrator switch array}
    \label{sec_td_switches}
    
        \subsubsection{Switch manufacturing and control electronics}
        \label{sec_TD_switch_manufacturing_and_control_electronics}
        
            The TD switch array design is based on the single channel prototype.
It reflects the 8$\times$8 structure of the feed-horn array which was chosen as a trade-off between the required filling factor and the possibility to leave enough room for the screws used to pack the platelets and connect the horns to the switch array.
It was designed at APC in Paris and manufactured at the machine-shop of the University of Milano-Bicocca using Al-6061T5 for the two main shells. 
The main body is made by two parts. The first is a base housing most of the \newb{waveguide}, the printed circuit board (PCB) and the coils+shutters. The bottom part of the base has the threaded holes to interface with the bottom horn array. The second part is a lid with the rest of the \newb{waveguide} length and the threaded holes to interface with the top horn array. 
The shutters are a replica (64 times) of the single channel shutter. They are mounted on a custom PCB sharing the same footprint of the \newb{waveguides}. Due to the limited cooling power at 1\,K, the number of wires reaching the 1\,K zone is reduced to a minimum. The heat dissipated by the coils while biased to close the shutters is also non negligible. So, in the 8$\times$8 TD it is possible to \new{close} only a small fraction of the shutters and not in an arbitrary configuration. 

 A modular electronics is used to operate the shutters and read their position.  Every electronic module is in charge of fifty shutters, being able to operate two of them at a time. Every module is composed of two arbitrary current generators to bias the coils. A small sinusoidal modulation of the coil current is used to read the shutter position, using the different phase delay between current and voltage when the ferrite is outside (shutter open) or inside (shutter closed) the coil. A set of six \new{Analog to Digital Converters} capture the alternate current (AC), AC voltage and direct current (DC) voltage during the 8\,ms after the command to close a shutter is set. The DC voltage is used to evaluate the ohmic load of the lines+coil system, in order to verify that there is electrical continuity. The AC signals are used to calculate the phase delay in a way similar to what is done in phase locked loop (PLL) circuits. The sinusoidal current and voltage signals are digitized as follows: when the signal is above its average value, it is recorded as 1, otherwise as 0. In this way two squared waves are generated, one for the voltage and one for the current, which phase delay is easily computed as $\pi$ times the average value of a third square wave obtained by applying XOR operator to the first two. When a switch is set to {\em closed}, a current pulse of 350\,mA lasting $\sim 5$ms is used to energize the switch coil and the ferrite is attracted inside it. Once the ferrite is inside, the current is reduced to 90\,mA, which is enough to hold in place to shutter reducing the heat load.

 An alternative method to compute the phase difference was tested too. It is based on the fits of the digitized signals (voltage and current) and the consequent computation of the phase delay. This approach, however, has a computational cost much higher than the XOR method because of the necessity to digitally filter the data and run fitting algorithms. A direct comparison of the two methods showed us that the two are perfectly compatible within the statistical uncertainties.

        \subsubsection{Switch cryogenic tests}
        \label{sec_TD_switch_cryogenic_test}
            
\label{switch_cryogenic_test}

We tested the TD switch array at 5\,K in the Milano Bicocca Millimeter Lab using a custom cryofacility.
The cryofacility is composed of an aluminum vacuum chamber operated by Cryomech PT407 pulse tube and equipped with a Varian Navigator turbo pump assisted by an Agilent Scroll roughing pump. A SRS CTC-100 cryogenic temperature controller is in charge of the temperature readings and stabilization by means of load resistors. Temperature sensors are calibrated DT-470 Lake Shore diodes. The base temperature with no heat load is 3.5\,K. 

\begin{figure}[h!]
    \begin{center}
        \includegraphics[width=12cm]{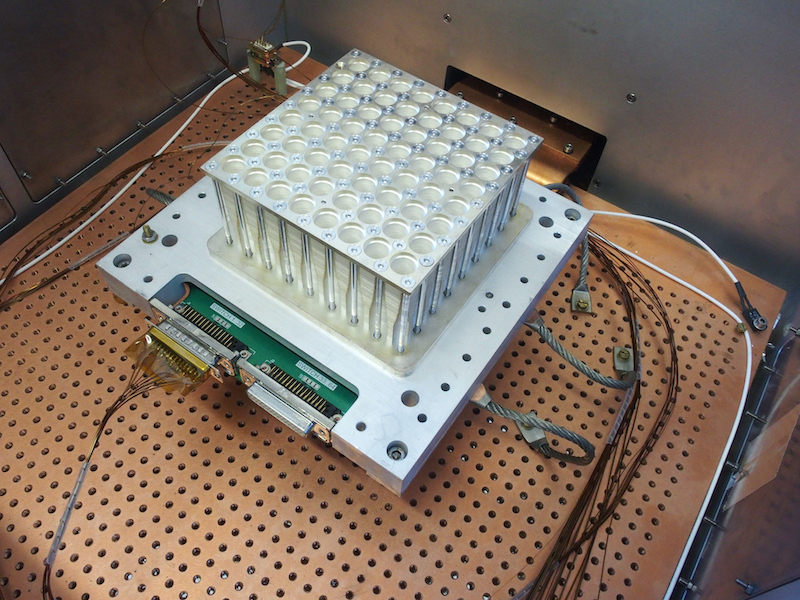}
        \caption{\label{fig_TD_sw_horns_inside_cryofacility}Technological demonstrator switch and horn array assembled and placed in the cryofacility for cryogenic functional tests.}
    \end{center}
\end{figure}

\newc{Once we assembled the TD switch array with the two horn arrays, we placed it in the cryofacility which reached 5\,K instead of 4\,K because of the heat load of the numerous needed harness wires.} We considered 5\,K to be representative of the behaviour of the device at its nominal working temperature (1\,K).
The aim of these tests was to find the effective current values to move the shutters and maintain them in closed position in a harness configuration similar to that in the QUBIC cryostat and with the shutter release spring stiffness due to cryogenic temperature, very similar to the operative one. Moreover, an important goal of the tests was to verify the switch functionality and reliability at cryogenic temperature.

\begin{figure}[h!]
    \begin{center}
        \includegraphics[width=12cm]{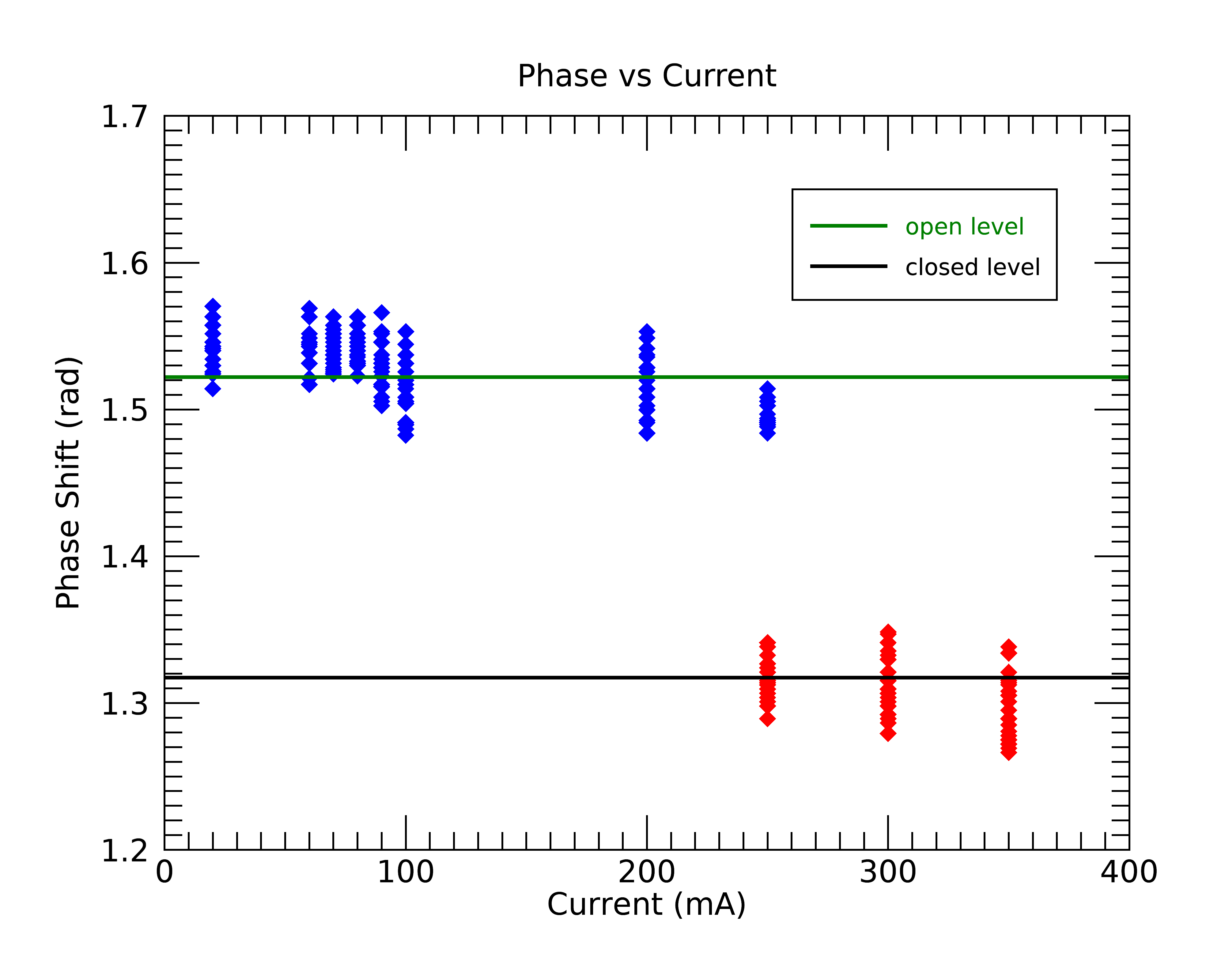}
        \caption{\label{fig_TD_current_pulse_oprimization_5K}Technological demonstrator switch excitation current optimization. All the switches move (and the phase changes) for current as high as 300\,mA. A conservative value of 350\,mA was adopted to assure the closing movement.}
    \end{center}
\end{figure}

\begin{figure}[h!]
    \begin{center}
        \includegraphics[width=12cm]{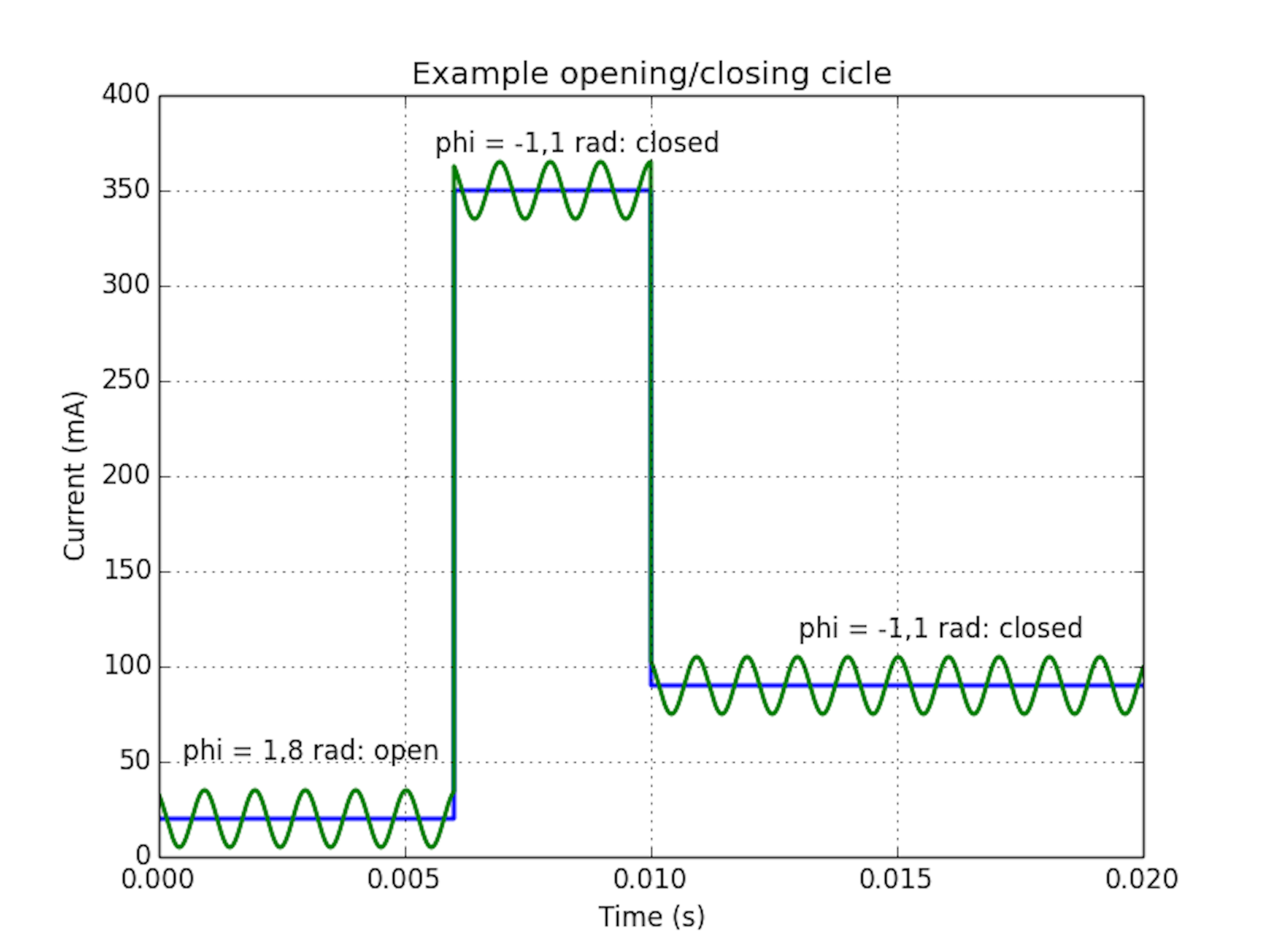}
        \caption{\label{fig_TD_driving_current_example_5K}Technological demonstrator switch excitation and readout current example. The excitation continuous current, plus the monitoring sinusoidal one, is kept at 20\,mA for the first 6\,ms, then a 350\,mA pulse lasting 4\,ms is applied to make the ferrite enter the coil and move the shutter. After 10\,ms the current is lowered to 90\,mA to reduce heat load.}
    \end{center}
\end{figure}

Figure~\ref{fig_TD_current_pulse_oprimization_5K} shows the phase between voltage and current for every switch and for increasing values of exciting current. 
We applied the excitation current from 20\,mA up to 350\,mA. All the switches did not move till a current value of 250\,mA, for which part of them closed, and the rest stayed open. At 300\,mA all the switches moved to closed position. We set a default conservative value of 350\,mA. We also found that once a switch is closed, a much lower holding current can be applied. Experimental verification led to a value of 90\,mA as the holding current capable to keep every switch in closed position.
A typical excitation pulse at cryogenic temperature with the superimposed oscillating current can be seen in figure~\ref{fig_TD_driving_current_example_5K}. The phase value doesn't change from 350\,mA to 90\,mA, proving that the switch is kept in closed position even with the lower current.

\begin{figure}[h!]
        \begin{center}
            \includegraphics[width=7.5cm]{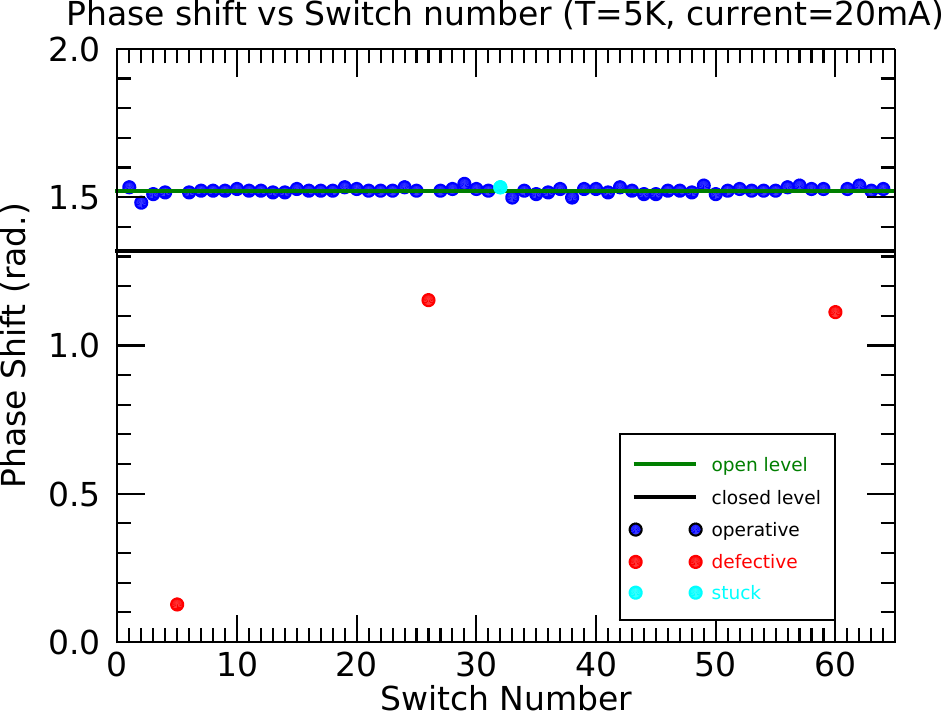}
            \includegraphics[width=7.5cm]{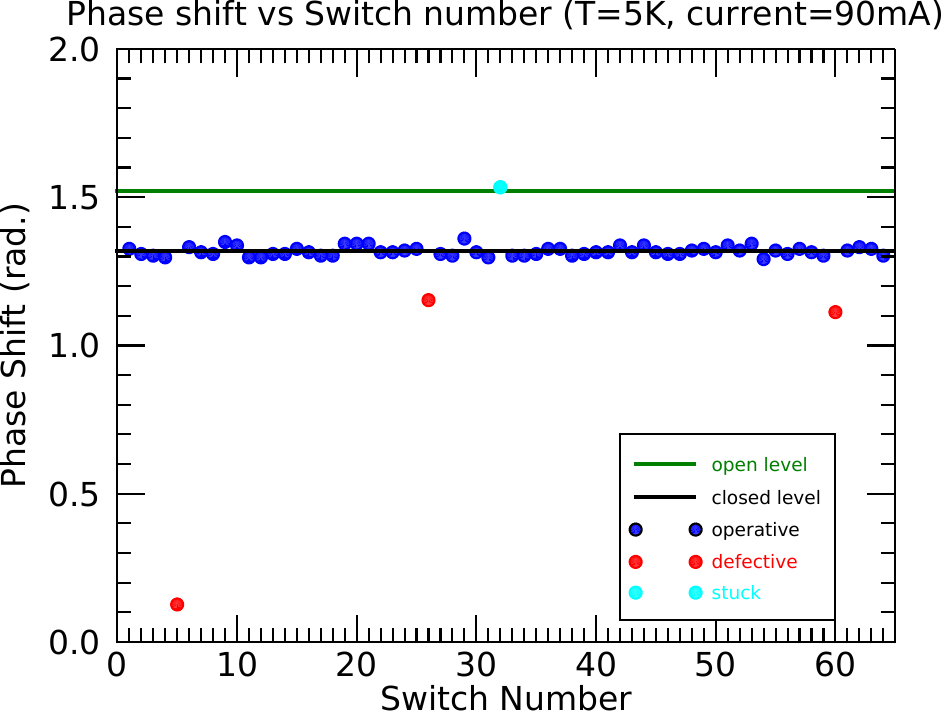}
        \end{center}
        \caption{\label{fig_TD_switch_operation} Phase between voltage and current measured by XOR algorithm.  \textit{Left}: shutters in open position \textit{Right}: shutters in closed position. Switch n.\,1 was operative at 5\,K. N.\,5, 26 and 60 were defective and n.\,32 was stuck open.}
    \end{figure}

Once we found the proper excitation current, we tested all the switches one by one, starting from room temperature. Three of them were defective (numbers 5, 26 and 60 because open or short-circuited) and two of them (n.\,1 and n.\,32) were mechanically stuck in the open position. These last two switches were stuck probably because of a non perfectly plane shutter. However, at cold temperature n.\,1 resulted operative. The result of this test is reported in figure~\ref{fig_TD_switch_operation}.

Finally we performed a statistical test operating each working switch 100 times. We computed the phase delay in open and closed position using both the functional fit and the XOR algorithm, applied to the data acquired by the AC coupled ADCs. Both methods gave consistent values, as reported in table~\ref{tab_switch_statistics}.

\begin{table}[ht]
\centering
\caption{Phase delay statistics in radians for 60 switches and 100 cycles each}
\label{tab_switch_statistics}
\begin{tabular}{|ll|ll|}
\hline
FIT                             &           & XOR                            &           \\ \hline
\multicolumn{1}{|l|}{Open}      & Closed    & \multicolumn{1}{l|}{Open}      & Closed    \\ \hline
\multicolumn{1}{|l|}{1.52$\pm$0.01} & 1.32$\pm$0.01 & \multicolumn{1}{l|}{1.52$\pm$0.01} & 1.33$\pm$0.01 \\ \hline
\end{tabular}
\end{table}

However, every switch shows a distinctive mean phase delay with a standard deviation which is one third of the one of the full population, reflecting on one side the intrinsic different impedance of each single coil, and on the other side the intrinsic accuracy of the phase measurement of our electronics which is of the order of 0.003 rad (10'). For this reason, we defined a standard calibration procedure to be run when the focal plane is cold, to compile a look-up table of the phase delay of every switch in open and closed position, based on a statistical analysis of 100 switch operations. The look-up table will then be used to decide if the switch moves in the right position or not and, if needed, to repeat the command.

Despite the overall good behaviour of the TD switch array, the mechanical tolerances needed to operate the shutters resulted in a certain level of unreliability. The main issue is
that a few shutters didn’t completely enter the wave-guide, even if their ferrites were totally inside the coil. \new{The main reason of this is the hook soldered on the ferrite responsible for the back and forth movement which was not sufficiently controlled to enable the ideal displacement of the shutter. The design of the stainless steel blade was consequently redesigned to reduce this uncertainty and improve the reliability of the switch motion}.

    \subsection{The full instrument switch array}
    \label{sec_fi_switches}
    The Full Instrument switch array was designed replicating the TD and adapting it to fill the circular aperture sampled by the 400 horns. \newe{Unfortunately, the TO5-shaped coils were out of stock and were no longer manufactured. Due to size constraints, selection of other switches was limited. We selected a bi-stable micro-shutter (the BOS7/10, manufactured by the Japanese company Takano Co. LTD) and modified the design to fit the volume available. The driving electronics that was developed for the previous shutters was readily adapted to the new devices. Because the new shutters are bi-stable they do not dissipate energy except for the few milliseconds when changing state.} They can be operated, in sequence, in any number from 1 to 400. Since this model is designed for room temperature operations, we tested it at 5\,K to confirm that it remains operative at cryogenic temperatures. We collected several thousands of movements without any problem, simply adjusting the excitation current with temperature.
The Full Instrument switch array \newe{is currently being developed}. 

 \begin{figure}[h!]
    \begin{center}
        \includegraphics[width=12cm]{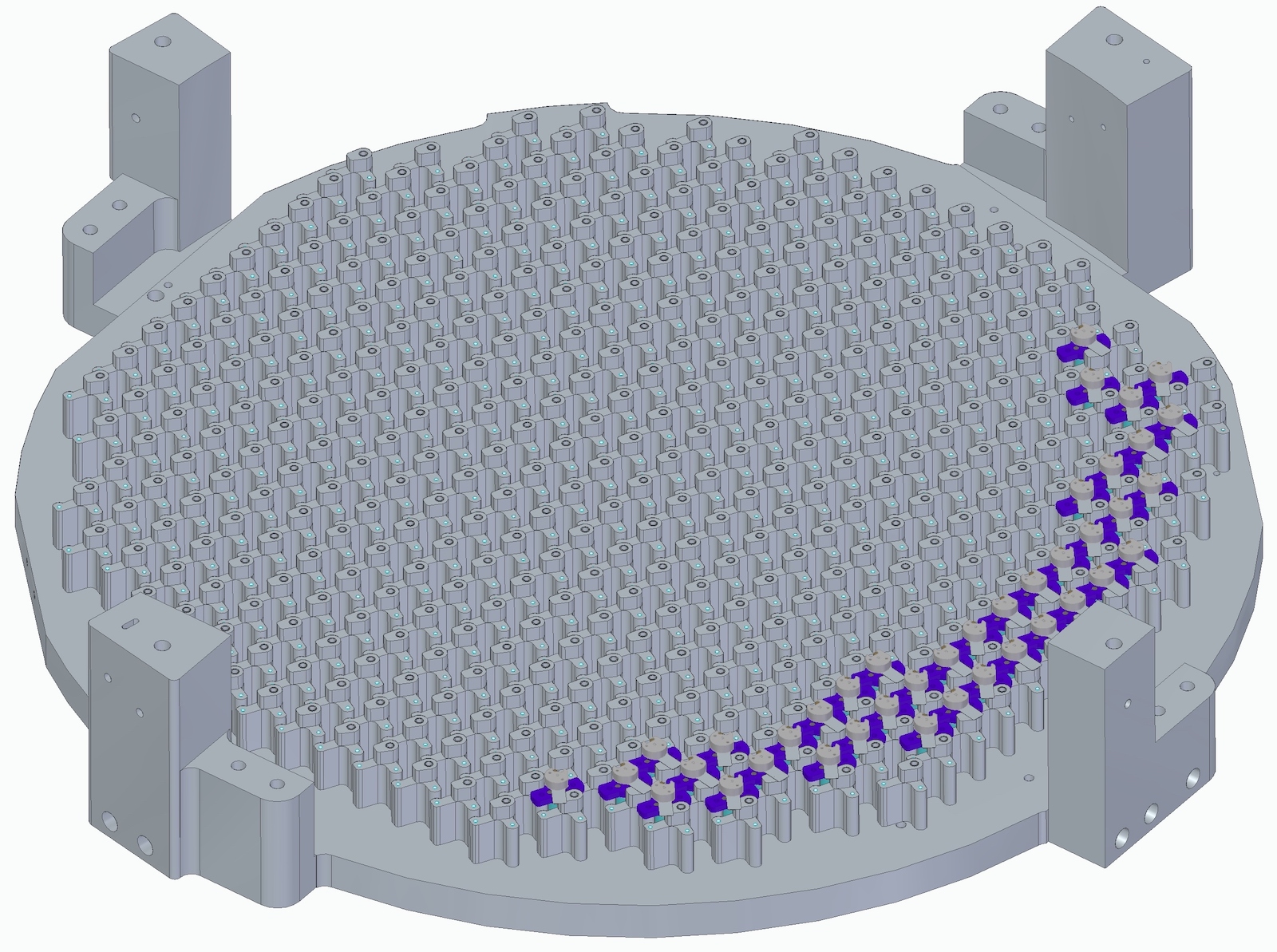}
        \caption{\label{fig_FI_switch_array} Open view of the Final Instrument switch array partially populated by the new bi-stable shutter.}
    \end{center}
\end{figure}

\section{Conclusions}
\label{sec_conclusions}

	In this paper we have described the design, manufacturing and testing of the feed-horn-switch system of the QUBIC technological demonstrator (TD), which will demonstrate the concept of bolometric interferometry by observing the polarized microwave sky from the high-altitude Alto Chorillo site in Argentina. The TD is the precursor of the QUBIC full instrument (FI), that will measure the CMB polarization from the same site.

The TD horn-switch system is composed by a square array of 64 back-to-back corrugated feed-horns interspersed by 64 mechanical switches that can open and close the connecting circular \newb{waveguides}. We designed the horns to allow the operation within two wide bands centered at 150 and 220\,GHz, and manufactured them in platelets that were drilled with a combination of photo-chemical etching (inner, 0.3\,mm plates) and mechanical milling (outer, 3\,mm and 6\,mm flanges).

We fully characterized the mechanical profile of all horns and found that the hole diameters of the inner plates were, on average, larger than the expected tolerance. The reason was the etching time that was not properly controlled during the process. We eliminated this problem in the full instrument (FI) horns and found that this out-of-spec was not critical for the objectives of the TD.

\newc{The measured electromagnetic performance agrees with simulations. In particular we obtained a return loss around $-$20\,dB up to 230\,GHz and beam patterns in agreement with single-mode simulations down to $-$30\,dB.}

Regarding the switches, we performed electromagnetic characterizations on the single channel prototype, finding return and insertion losses at 150\,GHz \newb{consistent} with expectations ($<-$25\,dB and $\sim-0.1$\,dB, respectively) and an isolation larger than 70\,dB (specification was 50\,dB). At 220\,GHz, return and insertion losses specifications are not met, but this is expected because the prototype was designed and manufactured for the QUBIC lower band, which is the only one operative in the TD. 
We successfully tested the switch array at room temperature and inside the QUBIC cryostat, at 5\,K, close to the nominal final one (1\,K). \newd{We developed a readout system able to monitor the actual switch positions in operative conditions (1\,K). Since the micro-miniaturized coils of the TD are not commercially available anymore, we found an alternative for the FI which forced us to redesign the whole mechanism around a room temperature commercial bi-stable shutter which has been already tested at 5\,K. This new shutter remained operative at cryogenic temperature simply adjusting the bias current.}

\newe{We have completed the development of the feed-horn arrays while the switch system is currently being developed}. After their mutual integration the whole system will be ready for the upgrade from the TD to the FI.

\section*{Acknowledgements}
\label{sec_acknowledgements}	
	
    QUBIC is funded by the following agencies. France: ANR (Agencie Nationale de la Recherche) 2012 and 2014, DIM-ACAV (Domaine d'Interet Majeur-Astronomie et Conditions d'Apparition de la Vie), Labex UnivEarthS (Universit\`e de Paris), CNRS/IN2P3 (Centre National de la Recherche Scientifique/Institut National de Physique Nucl\`eaire et de Physique des Particules), CNRS/INSU (Centre National de la Recherche Scientifique/Institut National des Sciences de l'Univers). Italy: CNR/PNRA (Consiglio Nazionale delle Ricerche/Programma Nazionale Ricerche in Antartide) until 2016, INFN (Istituto Nazionale di Fisica Nucleare) since 2017. Argentina: MINCyT (Ministerio de Ciencia, Tecnolog\'ia e Innovaci\'on), CNEA (Comisi\'on Nacional de Energ\'ia At\'omica), CONICET (Consejo Nacional de Investigaciones Cient\'ificas y T\'ecnicas). Ireland: D. Burke and J.D. Murphy acknowledge funding from the Irish Research Council under the Government of Ireland Postgraduate Scholarship Scheme. D. Gayer and S. Scully acknowledge funding from the National University of Ireland, Maynooth. D. Bennett acknowledges funding from Science Foundation Ireland.

\bibliographystyle{JHEP}
\bibliography{biblio,qubic}

\providecommand{\href}[2]{#2}\begingroup\raggedright\begin{thebibliography}{10}

\bibitem{2011APh....34..705Q}
{QUBIC Collaboration}, E.~{Battistelli}, A.~{Ba{\'u}}, D.~{Bennett},
  L.~{Berg{\'e}}, J.~P. {Bernard} et~al., \emph{{QUBIC: The QU bolometric
  interferometer for cosmology}},
  \href{https://doi.org/10.1016/j.astropartphys.2011.01.012}{\emph{Astroparticle
  Physics} {\bfseries 34} (2011) 705}
  [\href{https://arxiv.org/abs/1010.0645}{{\ttfamily 1010.0645}}].

\bibitem{2019Univ....5...42M}
A.~{Mennella}, P.~{Ade}, G.~{Amico}, D.~{Auguste}, J.~{Aumont}, S.~{Banfi}
  et~al., \emph{{QUBIC: Exploring the Primordial Universe with the Q\&U
  Bolometric Interferometer}},
  \href{https://doi.org/10.3390/universe5020042}{\emph{Universe} {\bfseries 5}
  (2019) 42}.

\bibitem{2020.QUBIC.PAPER1}
J.~C. {Hamilton}, L.~{Mousset} et~al., \emph{{QUBIC I: Overview and
  ScienceProgram}}, {\emph{arXiv e-prints} (2020) arXiv:2011.02213}
  [\href{https://arxiv.org/abs/2011.02213}{{\ttfamily 2011.02213}}].

\bibitem{2020.QUBIC.PAPER2}
L.~{Mousset}, M.~M. {Gamboa Lerena} et~al., \emph{{QUBIC II:
  Spectro-Polarimetry with Bolometric Interferometry}}, {\emph{arXiv e-prints}
  (2020) arXiv:2010.15119} [\href{https://arxiv.org/abs/2010.15119}{{\ttfamily
  2010.15119}}].

\bibitem{Bigot-Sazy2013}
M.-A. Bigot-Sazy, R.~Charlassier, J.~Hamilton, J.~Kaplan and G.~Zahariade,
  \emph{{Astrophysics Self-calibration : an efficient method to control
  systematic effects in bolometric interferometry}}, {\emph{Astronomy {\&}
  Astrophysics} {\bfseries 59} (2013) 1}.

\bibitem{haas1993}
R.~W. {Haas}, D.~{Brest}, H.~{Mueggenburg}, L.~{Lang} and D.~{Heimlich},
  \emph{Fabrication and performance of mmw and smmw platelet horn arrays},
  \href{https://doi.org/10.1007/BF02085698}{\emph{International Journal of
  Infrared and Millimeter Waves} {\bfseries 14} (1993) 2289}.

\bibitem{haas1995}
R.~W. {Haas}, \emph{Further development of mmw and smmw platelet feed horn
  arrays},  in \emph{Multi-Feed Systems for Radio Telescopes}, D.~T. {Emerson}
  and J.~M. {Payne}, eds., vol.~75 of \emph{Astronomical Society of the Pacific
  Conference Series}, p.~99, January, 1995.

\bibitem{bock2009}
J.~J. {Bock}, J.~{Gundersen}, A.~T. {Lee}, P.~L. {Richards} and E.~{Wollack},
  \emph{Optical coupling},  in \emph{Journal of Physics Conference Series},
  vol.~155 of \emph{Journal of Physics Conference Series}, p.~012005, March,
  2009, \href{https://doi.org/10.1088/1742-6596/155/1/012005}{DOI}.

\bibitem{britton2010}
J.~W. {Britton}, J.~P. {Nibarger}, K.~W. {Yoon}, J.~A. {Beall}, D.~{Becker},
  H.-M. {Cho} et~al., \emph{Corrugated silicon platelet feed horn array for cmb
  polarimetry at 150 ghz},  in \emph{Millimeter, Submillimeter, and
  Far-Infrared Detectors and Instrumentation for Astronomy V}, W.~S. {Holland}
  and J.~{Zmuidzinas}, eds., vol.~7741 of \emph{Society of Photo-Optical
  Instrumentation Engineers (SPIE) Conference Series}, p.~77410T, July, 2010,
  \href{https://doi.org/10.1117/12.857701}{DOI}.

\bibitem{deltorto2011}
F.~{Del Torto}, M.~{Bersanelli}, F.~{Cavaliere}, A.~{De Rosa},
  O.~{D'Arcangelo}, C.~{Franceschet} et~al., \emph{W-band prototype of platelet
  feed-horn array for cmb polarisation measurements},
  \href{https://doi.org/10.1088/1748-0221/6/06/P06009}{\emph{Journal of
  Instrumentation} {\bfseries 6} (2011) 6009}
  [\href{https://arxiv.org/abs/1107.1157}{{\ttfamily 1107.1157}}].

\bibitem{deltorto2015}
F.~{Del Torto}, C.~Franceschet, F.~Villa, P.~Battaglia, M.~Bersanelli,
  F.~Cavaliere et~al., \emph{{A 7X7 elements Q-band feed-horn array for the
  LSPE-STRIP instrument}},  in \emph{36th ESA Antenna Workshop on Antennas and
  RF Systems for Space Science}, 2015.

\bibitem{2012JLTP..167..917N}
M.~D. {Niemack}, J.~{Beall}, D.~{Becker}, H.~M. {Cho}, A.~{Fox}, G.~{Hilton}
  et~al., \emph{Optimizing feedhorn-coupled tes polarimeters for balloon and
  space-based cmb observations},
  \href{https://doi.org/10.1007/s10909-012-0554-2}{\emph{Journal of Low
  Temperature Physics} {\bfseries 167} (2012) 917}.

\bibitem{bischoff2013}
C.~{Bischoff}, A.~{Brizius}, I.~{Buder}, Y.~{Chinone}, K.~{Cleary}, R.~N.
  {Dumoulin} et~al., \emph{The q/u imaging experiment instrument},
  \href{https://doi.org/10.1088/0004-637X/768/1/9}{\emph{Atrophysical Journal}
  {\bfseries 768} (2013) 9} [\href{https://arxiv.org/abs/1207.5562}{{\ttfamily
  1207.5562}}].

\bibitem{mandelli2021}
S.~Mandelli, E.~Manzan, A.~Mennella, F.~Cavaliere, D.~Vigan{\`{o}},
  C.~Franceschet et~al., \emph{{A chemically etched corrugated feedhorn array
  for D-band CMB observations}},
  \href{https://doi.org/10.1007/s10686-021-09698-9}{\emph{Experimental
  Astronomy} (2021) 1}.

\bibitem{2020.QUBIC.PAPER5}
S.~{Masi} et~al., \emph{{QUBIC V: Cryogenic system design and performance}},
  {\emph{arXiv e-prints} (2020) arXiv:2008.10659}
  [\href{https://arxiv.org/abs/2008.10659}{{\ttfamily 2008.10659}}].

\bibitem{2020.QUBIC.PAPER6}
G.~{D'Alessandro}, L.~{Mele}, F.~{Columbro} et~al., \emph{{QUBIC VI: cryogenic
  half wave platerotator, design and performances}}, {\emph{arXiv e-prints}
  (2020) arXiv:2008.10667} [\href{https://arxiv.org/abs/2008.10667}{{\ttfamily
  2008.10667}}].

\bibitem{2020.QUBIC.PAPER3}
S.~A. {Torchinsky} et~al., \emph{{QUBIC III: Laboratory Characterization}},
  {\emph{arXiv e-prints} (2020) arXiv:2008.10056}
  [\href{https://arxiv.org/abs/2008.10056}{{\ttfamily 2008.10056}}].

\bibitem{2020arXiv200810056T}
S.~A. {Torchinsky}, J.~C. {Hamilton}, M.~{Piat}, E.~S. {Battistelli},
  C.~{Chapron}, G.~{D'Alessandro} et~al., \emph{{QUBIC III: Laboratory
  Characterization}}, {\emph{arXiv e-prints} (2020) arXiv:2008.10056}
  [\href{https://arxiv.org/abs/2008.10056}{{\ttfamily 2008.10056}}].

\bibitem{2020.QUBIC.PAPER8}
C.~{O'Sullivan}, M.~{De Petris} et~al., \emph{{QUBIC VIII: Optical design and
  performance}}, {\emph{arXiv e-prints} (2020) arXiv:2008.10119}
  [\href{https://arxiv.org/abs/2008.10119}{{\ttfamily 2008.10119}}].

\bibitem{2020.QUBIC.PAPER4}
M.~{Piat}, G.~{Stankowiak} et~al., \emph{{QUBIC IV: Performance of TES
  Bolometers and Readout Electronics}}, {\emph{arXiv e-prints} (2021)
  arXiv:2101.06787} [\href{https://arxiv.org/abs/2101.06787}{{\ttfamily
  2101.06787}}].

\bibitem{maffei2016}
B.~Maffei, E.~Gleeson, J.~A. Murphy and G.~Pisano, \emph{{Study of corrugated
  Winston horns}},  in \emph{Millimeter and Submillimeter Detectors for
  Astronomy II}, J.~Zmuidzinas, W.~S. Holland and S.~Withington, eds.,
  vol.~5498, pp.~812 -- 817, International Society for Optics and Photonics,
  SPIE, 2004, \href{https://doi.org/10.1117/12.552922}{DOI}.

\bibitem{Clover2008}
L.~{Piccirillo}, P.~{Ade}, M.~D. {Audley}, C.~{Baines}, R.~{Battye}, M.~{Brown}
  et~al., \emph{{The ClOVER experiment}},  in \emph{Millimeter and
  Submillimeter Detectors and Instrumentation for Astronomy IV}, W.~D.
  {Duncan}, W.~S. {Holland}, S.~{Withington} and J.~{Zmuidzinas}, eds.,
  vol.~7020 of \emph{Society of Photo-Optical Instrumentation Engineers (SPIE)
  Conference Series}, p.~70201E, July, 2008,
  \href{https://doi.org/10.1117/12.788927}{DOI}.

\bibitem{quijote2010}
A.~{Gomez}, G.~{Murga}, B.~{Etxeita}, R.~{Sanquirce}, R.~{Rebolo}, J.~A.
  {Rubi{\~n}o-Martin} et~al., \emph{{QUIJOTE telescope design and
  fabrication}},  in \emph{Ground-based and Airborne Telescopes III}, L.~M.
  {Stepp}, R.~{Gilmozzi} and H.~J. {Hall}, eds., vol.~7733 of \emph{Society of
  Photo-Optical Instrumentation Engineers (SPIE) Conference Series}, p.~77330Z,
  July, 2010, \href{https://doi.org/10.1117/12.857286}{DOI}.

\bibitem{BINGO2020}
C.~A. {Wuensche}, L.~{Reitano}, M.~W. {Peel}, I.~W.~A. {Browne}, B.~{Maffei},
  E.~{Abdalla} et~al., \emph{{Baryon acoustic oscillations from Integrated
  Neutral Gas Observations: Broadband corrugated horn construction and
  testing}},
  \href{https://doi.org/10.1007/s10686-020-09666-9}{\emph{Experimental
  Astronomy} {\bfseries 50} (2020) 125}
  [\href{https://arxiv.org/abs/1911.13188}{{\ttfamily 1911.13188}}].

\bibitem{bicep2-2014}
{BICEP2 Collaboration}, P.~A.~R. {Ade}, R.~W. {Aikin}, D.~{Barkats}, S.~J.
  {Benton}, C.~A. {Bischoff} et~al., \emph{Detection of b-mode polarization at
  degree angular scales by bicep2},
  \href{https://doi.org/10.1103/PhysRevLett.112.241101}{\emph{Phys. Rev. Lett.}
  {\bfseries 112} (2014) 241101}
  [\href{https://arxiv.org/abs/1403.3985}{{\ttfamily 1403.3985}}].

\bibitem{james1981}
G.~L. {James}, \emph{Analysis and design of te/11/-to-he/11/ corrugated
  cylindrical waveguide mode converters},
  \href{https://doi.org/10.1109/TMTT.1981.1130499}{\emph{IEEE Transactions on
  Microwave Theory Techniques} {\bfseries 29} (1981) 1059}.

\bibitem{clarricoats}
P.~{Clarricoats} and A.~{Olver}, \emph{{Corrugated Horns for Microwave
  Antennas}}. Peter Peregrinus Ltd, 1984.

\bibitem{olver}
A.~{Olver}, P.~{Clarricoats}, A.~{Kishk} and L.~{Shafai}, \emph{{Microwave
  horns and feeds}}. IEEE Press, 1994.

\bibitem{murphy1991}
J.~Murphy and R.~Padman, \emph{Radiation patterns of few-moded horns and
  condensing lightpipes},
  \href{https://doi.org/https://doi.org/10.1016/0020-0891(91)90060-S}{\emph{Infrared
  Physics} {\bfseries 31} (1991) 291}.

\bibitem{murphy2001}
J.~Murphy, R.~Colgan, C.~O’Sullivan, B.~Maffei and P.~Ade, \emph{Radiation
  patterns of multi-moded corrugated horns for far-ir space applications},
  \href{https://doi.org/https://doi.org/10.1016/S1350-4495(01)00063-9}{\emph{Infrared
  Physics \& Technology} {\bfseries 42} (2001) 515}.

\bibitem{Kerr2009}
A.~R. {Kerr} and S.~{Srikanth}, \emph{{The Ring-CenteredWaveguide Flange for
  Submillimeter Wavelengths}},  in \emph{20th International Symposium on Space
  Terahertz Technology}, 2009.

\bibitem{SRSR}
A.~{Berthon} and R.~P. {Bills}, \emph{Integral equation analysis of radiating
  structures of revolution}, \href{https://doi.org/10.1109/8.18702}{\emph{IEEE
  Transactions on Antennas and Propagation} {\bfseries 37} (1989) 159}.

\bibitem{CLARK1970295}
A.~Clark, G.~Childs and G.~Wallace, \emph{Electrical resistivity of some
  engineering alloys at low temperatures},
  \href{https://doi.org/https://doi.org/10.1016/0011-2275(70)90056-1}{\emph{Cryogenics}
  {\bfseries 10} (1970) 295}.

\bibitem{Handbook...Al...1990}
\emph{ASM HANDBOOK Properties and Selection: Nonferrous Alloys and
  Special-Purpose Materials}. ASM International, 1990,
  \href{https://doi.org/DOI: https://doi.org/10.31399/asm.hb.v02.a0001060}{DOI:
  https://doi.org/10.31399/asm.hb.v02.a0001060}.

\bibitem{Bordier2014}
G.~Bordier, \emph{{D\'eveloppements de composants millim\'etriques pour la
  caract\'erisation de la polarisation du fond diffus cosmologique}}, phd,
  \`Ecole doctorale ED127: Astronomie \& Astrophysique d'\^Ile de France -
  Laboratoire Astroparticule et Cosmologie (APC), 2014.

\bibitem{1949JAP....20..352M}
S.~P. {Morgan}, \emph{{Effect of Surface Roughness on Eddy Current Losses at
  Microwave Frequencies}},
  \href{https://doi.org/10.1063/1.1698368}{\emph{Journal of Applied Physics}
  {\bfseries 20} (1949) 352}.

\bibitem{1992IMGWL...2..180C}
C.-D. {Chen}, C.-K.~C. {Tzuang} and S.~T. {Peng}, \emph{{Full-wave analysis of
  a lossy rectangular waveguide containing rough inner surfaces}}, {\emph{IEEE
  Microwave and Guided Wave Letters} {\bfseries 2} (1992) 180}.

\bibitem{2012SPIE.8452E..2RZ}
M.~{Zannoni}, A.~{Ba\`u}, A.~{Passerini}, A.~{Tartari}, M.~{Gervasi} and
  L.~{Valenziano}, \emph{{A cryogenic set-up for accurate measurements of
  S-parameters}},  in \emph{Millimeter, Submillimeter, and Far-Infrared
  Detectors and Instrumentation for Astronomy VI}, W.~S. {Holland} and
  J.~{Zmuidzinas}, eds., vol.~8452 of \emph{Society of Photo-Optical
  Instrumentation Engineers (SPIE) Conference Series}, p.~84522R, Sept., 2012,
  \href{https://doi.org/10.1117/12.925995}{DOI}.

\end{thebibliography}\endgroup

\end{document}